\documentclass[reprint,amsmath,amssymb,aps,nofootinbib]{revtex4-2}

\usepackage[utf8]{inputenc}
\usepackage[T1]{fontenc}
\usepackage{lmodern}
\usepackage[english]{babel}

\usepackage{amssymb,amsmath,amsfonts}
\usepackage{graphicx}
\usepackage{xparse}
\usepackage{xstring}
\usepackage{xcolor}
\usepackage{booktabs}
\usepackage{paralist}
\usepackage{enumitem}
\usepackage{csquotes}
\usepackage{nicefrac}

\usepackage{nameref}
\usepackage{hyperref}
\usepackage[capitalise]{cleveref} 
\usepackage{url} 

\usepackage{tabularray}
\UseTblrLibrary{booktabs,siunitx}

\input{aas_macros.sty}

\def \e         {\mathrm{e}} 
\def \d         {\mathrm{d}} 

\DeclareMathOperator{\diag}{diag}

\newcommand \brackets[2][]{\left[\vphantom{#1}#2\right]} 
\newcommand \parantheses[2][]{\left(\vphantom{#1}#2\right)} 
\newcommand \braces[2][]{\left\{\vphantom{#1}#2\right\}} 

\def \diff          #1#2{\frac{\d#1}{\d#2}}

\def \tdiff          #1#2{\d#1/\d#2}

\renewcommand{\vec}[1]{\mbox{\boldmath$#1$}}

\hypersetup{
	colorlinks,
	linkcolor={red!50!black},
	citecolor={magenta},
	urlcolor={blue!80!black}
}

\renewcommand{\arraystretch}{1.2}
\SetTblrDefault{rowsep=3pt, colsep=1pt}

\DeclareSIUnit \speedoflight {c}
\DeclareSIUnit \parsec {pc}
\DeclareSIUnit \eV {eV}
\DeclareSIUnit \eVcc {eV/$c^2$}
\DeclareSIUnit \Msun {M_\odot}

\setlist[description]{
  leftmargin=\parindent
}

\NewDocumentCommand{\DEFradius}{O{1}}{%
	\ifnum #1>0 %
		\frac{r^{#1}}{R^{#1}} %
	\else %
		\ifnum #1=-1 %
			\frac{R}{r} %
		\else %
			\frac{R^{\StrGobbleLeft{#1}{1}}}{r^{\StrGobbleLeft{#1}{1}}} %
		\fi %
	\fi %
}

\newcommand \SCLradiusB {R_{\text{\tiny B}}}

\NewDocumentCommand{\DEFradiusB}{O{1}}{%
	\ifnum #1>0 %
		\frac{r^{#1}}{\SCLradiusB^{#1}} %
	\else %
		\ifnum #1=-1 %
			\frac{\SCLradiusB}{r} %
		\else %
			\frac{\SCLradiusB^{\StrGobbleLeft{#1}{1}}}{r^{\StrGobbleLeft{#1}{1}}} %
		\fi %
	\fi %
}

\def \SCLfermiEnergy{\varepsilon_{\text{\tiny F}}}

\NewDocumentCommand{\DEFfermiEnergy}{O{1}}{%
	\ifnum #1=1 %
		\SCLfermiEnergy(r) %
	\else %
		\SCLfermiEnergy^{#1}(r) %
	\fi %
}

\def \SYMmassDM {M_{\rm DM}}

\begin{document}

\title[Thermodynamics of self-gravitating fermions as a robust theory for dark matter halos: Stability analysis applied to the Milky Way]{
    Thermodynamics of self-gravitating fermions as a robust theory for dark matter halos:\texorpdfstring{\\}{}
    Stability analysis applied to the Milky Way
}

\author{A. Krut}
\affiliation{ICRANet, Piazza della Repubblica 10, I-65122 Pescara, Italy}

\author{C. R. Argüelles}
\affiliation{Instituto de Astrof\'isica de La Plata, UNLP-CONICET, Paseo del Bosque s/n B1900FWA La Plata, Argentina}
\affiliation{ICRANet, Piazza della Repubblica 10, I-65122 Pescara, Italy}
\email{carguelles@fcaglp.unlp.edu.ar}

\author{P.-H. Chavanis}
\affiliation{Laboratoire de Physique Th\'eorique, Universit\'e de Toulouse, CNRS, UPS, France}

\date{\today}


\begin{abstract}
We present a framework for dark matter halo formation based on a kinetic theory of self-gravitating fermions together with a solid connection to thermodynamics. Based on maximum entropy arguments, this approach predicts a most likely phase-space distribution which takes into account the Pauli exclusion principle, relativistic effects, and particle evaporation. The most general equilibrium configurations depend on the particle mass and develop a degenerate compact core embedded in a diluted halo, both linked by their fermionic nature. By applying such a theory to the Milky Way we analyze the stability of different families of equilibrium solutions with implications on the dark matter distribution and the mass of the dark matter particle candidate. We find that stable \textit{core}--\textit{halo} profiles, which explain the dark matter distribution in the Galaxy, exist only in the range $mc^2 \approx \SIrange{194}{387}{\kilo\eV}$. The lower bound is a consequence of imposing thermodynamical stability on the \textit{core}--\textit{halo} solutions having a $\SI{4.2E6}{\Msun}$ quantum core mass alternative to the black hole hypothesis at the Galaxy center. The upper bound is solely an outcome of general relativity when the quantum core reaches the Oppenheimer-Volkoff limit and undergoes gravitational collapse towards a black hole.  We demonstrate that stable \textit{core}--\textit{halo} profiles exist which are astrophysically relevant in the sense that their total mass is finite, do not suffer from the gravothermal catastrophe, and agree with observations. The morphology of the halo tail is described by a polytrope of index $\nicefrac{5}{2}$, developing a sharp decline of the density beyond $\SI{25}{\kilo\parsec}$ in excellent agreement with the latest Gaia DR3 rotation curve data. Moreover, we obtain a total mass of about $\SI{2E11}{\Msun}$ including baryons and a local dark matter density of about $\SI{0.4}{\giga\eV \per \speedoflight\squared \per \centi\metre\cubed}$ in line with recent independent estimates.
\end{abstract}

\maketitle


\section{Introduction}
\label{sec:introduction}

Current attempts to explain the process towards relaxation of dark matter (DM) halos and their final quasi-universal density profiles are mainly centered in $N$-body simulations (see \cite{2012AnP...524..507F} for a review). Despite their success in providing a fitting formula or law for the virialized DM distributions composed of collisionless particles --- from galactic to cluster scales and on different cosmological epochs \citep{1997ApJ...490..493N,1997ApJS..111...73K,2008Natur.454..735D,2009MNRAS.398L..21S,2010MNRAS.402...21N} --- we still lack a clear understanding on its physical basis. Considerable efforts have been performed in the past to try to explain the shape of the DM halos in terms of first principle physics such as statistical mechanics and thermodynamics. First attempts (prior to numerical analyses) were applied to idealized galaxy systems using either `phase-cells' \citep{1957SvA.....1..748O,1967MNRAS.136..101L} or particle statistical physics methods \citep{1978ApJ...225...83S,1987ApJ...316..497M} aimed to find a most probable distribution function at relaxation. 

In the pioneering work of Lynden-Bell \cite{1967MNRAS.136..101L} it was proposed a relaxation mechanism for collisionless systems which may explain the relaxed velocity distributions observed in galaxies in a fraction of the age of the Universe, a matter which was long known not to be feasible within traditional collisional relaxation processes driven by purely gravitational encounters \citep{Chandrasekhar1942,Binney2008}. The Lynden-Bell collisionless approach to equilibrium, called `violent relaxation', assumes a maximum entropy principle (MEP) as the responsible of driving such a self-gravitating system towards the observed steady state. However, on maximizing the entropy at fixed total mass and total energy, Lynden-Bell faced serious problems \citep{1990PhR...188..285P}: (i) he found a density profile decreasing as $\rho(r) \sim r^{-2}$ at large distances, so that the equilibrium state has an infinite mass (in contradiction with the initial assumption), meaning that there is no entropy maximum at fixed mass and energy; (ii) in the nondegenerate (dilute) limit of his theory (which may be particularity relevant to classical particles) the system undergoes a form of collisionless `gravothermal catastrophe' --- it occurs an entropy runaway by transferring energy from inner to outermost orbits without bound ---  and (iii) the resulting equilibrium distribution disagrees with those obtained from numerical simulations and observations. Therefore, Lynden-Bell \cite{1967MNRAS.136..101L} concluded that violent relaxation of self-gravitating systems must be incomplete.

Interestingly, the above difficulties faced by those early attempts on violent relaxation have been mostly superseded in the last decade. As already recognized in \cite{1990PhR...188..285P} --- given that the collisionless evolution of the system conserves an infinite set of quantities --- in \cite{2013MNRAS.430..121P} a MEP was applied though this time incorporating the conservation of a radial action during the rapid changes of the gravitational potential. This new dynamical constraint on orbital actions provides a different solution from that originally found in \cite{1967MNRAS.136..101L} based on mass and energy conservation alone, further accounting for the incompleteness of the violent relaxation. The main outcome of this new MEP approach is to produce a good match with $N$-body simulations in the outer DM density profile, with $\rho \sim r^{-3}$, albeit some discrepancy with CDM-only simulations remained towards the inner halo leaving still open a clear physical interpretation of the central $\rho \sim r^{-1}$ cusp \citep{2013MNRAS.430..121P}.

Further progress based on MEP approaches to find equilibrium profiles from statistical mechanics were advanced in \cite{2015ApJ...811....2H,2022ApJ...937...67W} based on incomplete relaxation, reaching a better overall agreement with numerical DM-only simulations. However, as recognized in those works, such statistical methods fall short for a full theory of DM halo formation and would need a kinetic theory framework together with a solid connection to thermodynamics.         

In this paper we propose a possible realization of such a full theory while further incorporating the quantum nature of the self-gravitating particles. Our approach is built upon a kinetic theory for massive neutral fermions taking tidal effects into account while it connects with a thermodynamical stability analysis, and at the same time does not suffer from any of the problems (i), (ii), (iii) stated above. That is, the fermionic DM profiles resulting from our theory have a finite mass, do not suffer from the gravothermal catastrophe, correspond to a true statistical equilibrium state (maximum entropy state), and agree with observations.

Our original line of research is based on a series of former theoretical works in the field \citep{1998MNRAS.300..981C,2006IJMPB..20.3113C,2015MNRAS.451..622R,2015PhRvD..91f3531C,2015PhRvD..92l3527C,2019PhRvD.100l3506C,2021MNRAS.502.4227A}, whose phenomenological analysis using modern data coming from different galaxy types including our own, was recently shown in \cite{2018PDU....21...82A,2019PDU....24..278A,2020A&A...641A..34B,2023ApJ...945....1K}. See \cite{2023mgm..conf.2230C,2023Univ....9..197A} for a historical account of the topic and an exhaustive list of references.

In the following, we highlight the relevance and benefits of applying a MEP approach and a thermodynamical analysis when compared with traditional simulation studies. Having a kinetic theory for self-gravitating particles allows for a self-consistent way to calculate analytic formulae for a most-likely distribution function (DF) accounting for particle evaporation, which is asked to maximize an entropy functional at relaxation (as done in this paper). Once with an expression for such a DF it allows for a semi-analytical treatment to calculate DM density profiles from equilibrium equations which can be integrated from the very center to the periphery, reaching a radial coverage not feasible within $N$-body simulations and being in excellent agreement with observations (see e.g. \cite{2023ApJ...945....1K} for recent results). 

The resulting equilibrium DM profiles can develop density tails in good agreement with simulations (i.e. $\rho \sim r^{-3}$), though depending on the degree of concentration of particles on halo scales they can be of polytropic shape \citep{2015PhRvD..91f3531C,2023ApJ...945....1K}. Interestingly, the inner-intermediate halo morphology of such DM profiles resulting from our MEP approach develops --- just after the quantum core --- a plateau resembling the Burkert (or cored Einasto) profiles, thus not suffering from the core-cusp tension associated with CDM-only profiles (see \cite{2023ApJ...945....1K} for an explicit comparison). Indeed, as claimed in \cite{2021MNRAS.504.2832S} from an independent MEP approach as the one applied here, the presence of DM cores on intermediate halo scales can be used as evidence of thermodynamic equilibrium, thus offering an alternative insight to the problem of DM halo formation.      

Moreover, our theoretical framework further allows us to incorporate the quantum nature of the DM particles with important physical and astrophysical consequences. As it will be shown here, from the physical point of view, it implies a source of quantum pressure (arising from the Pauli exclusion principle) supporting the stability of the innermost halo region while avoiding the gravothermal catastrophe. The astrophysical role of such a high density quantum-core arising at the center of the DM halo has been studied for both, bosonic DM (see e.g. \cite{2019PhRvD.100h3022C,2022EPJB...95...48C} for a thermodynamical semi-analytical treatment and \cite{2014NatPh..10..496S} for a wave simulation approach) and fermionic DM (see e.g. \cite{2015PhRvD..92l3527C,2020EPJB...93..208A,2021MNRAS.502.4227A,2022PhRvD.106d3538C} for thermodynamical semi-analytical approaches). The latter holds deep implications to the possible nature of compact objects (like SgrA*) at the center of the galaxies \citep{2020A&A...641A..34B,2021MNRAS.505L..64B,2022MNRAS.511L..35A,2024MNRAS.534.1217P}, and to the problem of supermassive black hole (SMBH) formation and growth in the early Universe \citep{2023MNRAS.523.2209A,2024ApJ...961L..10A}. 

In this work we give special attention to the Milky Way (MW). Our formation and stability study is performed in a fully general relativistic framework following recent results applied to average size galaxies in \citep{2021MNRAS.502.4227A} . We explicitly calculate the possible family of \textit{dense core} -- \textit{diluted halo} fermionic profiles predicted by this theory and aimed to explain the Grand Rotation Curve of the Galaxy --- i.e. with data ranging from Galaxy center to outer halo \citep{2013PASJ...65..118S} --- while further including for recent GAIA DR3 data \citep{2023A&A...678A.208J}.  In particular, we focus on thermodynamical stability criterion for DM halos. Thus, we study the thermodynamical stability of fermionic core-halo solutions within a formation scenario for DM halos based on a MEP. We locate the position of these solutions on the caloric curve of general relativistic self-gravitating fermions and determine the range of the DM particle mass $m$ for which these solutions fall on the stable branch. We show that there exists a set of stable and astrophysical core-halo solutions such that the DM halo is in good agreement with recent GAIA DR3 rotation curve data, while the dense DM core (fermion ball) works as an alternative to the central black hole hypothesis. In this manner we can reproduce the Grand Rotation Curve of the Galaxy from the inner region to halo scales with a single distribution function for the DM component. Hence, we recover previous results where the dense quantum core mimics the compact object SgrA* at the Galaxy center \citep{2020A&A...641A..34B,2021MNRAS.505L..64B,2022MNRAS.511L..35A,2024MNRAS.534.1217P}, though for the first time we analyze here the thermodynamic stability of the solutions, further restricting the particle mass range.


\section{Theory}
\label{sec:theory}
We introduce the system of equations for a self-gravitating system of massive neutral fermions in hydrostatic equilibrium within general relativity. More specifically, we consider a spherically symmetric system for a perfect fluid whose equation of state (EoS) takes into account (i) the quantum and relativistic effects of the fermionic particles, (ii) finite temperature effects of the system, and (iii) escape of particles effects (i.e. evaporation) through a cutoff in the Fermi–Dirac distribution function (DF).

Before we provide the necessary equations describing the problem of fermionic mass distribution, we introduce proper scaling factors to eliminate the particle mass $m$ (and other physical constants) from the equations. In particular, we introduce scaling factors for mass, length and density labeled as $M$, $R$ and $\rho$, respectively. We follow here a notation closer to the mathematical convention of distinguishing between the symbol, e.g., $\rho$ for a constant and the notation $\rho(r)$ for a function.

It is convenient to describe them in the Planck unit system covering the gravitational constant $G$, the reduced Planck constant $\hbar$ and the speed of light $c$. With the Planck mass $m_{\rm P} = \sqrt{\hbar c / G}$, the Planck length $l_{\rm P} = \sqrt{\hbar G / c^3}$ and the Planck density $\rho_{\rm P} = m_{\rm P} / l_{\rm P}^3$ we obtain a unit system depending only on the particle mass $m$ and the particle degeneracy $g$. They are given by \begin{align}
    \label{eqn:scale:mass}
    \frac{M}{m_{\rm P}}       & = \frac12 g^{-1/2} \pi^{1/4} \brackets{\frac{m}{m_{\rm P}}}^{-2}, \\
     \label{eqn:scale:length}
    \frac{R}{l_{\rm P}}       & = g^{-1/2} \pi^{1/4} \brackets{\frac{m}{m_{\rm P}}}^{-2}, \\
     \label{eqn:scale:density}
    \frac{\rho}{\rho_{\rm P}} & = \frac{1}{8} g \pi^{-3/2} \brackets{\frac{m}{m_{\rm P}}}^{4}.
\end{align} Here, the scale factors are related through $R = 2GM / c^2$ and $M = 4\pi R^3 \rho$. For fermions we use $g = 2$. It is worth mentioning that these scaling factors are of the order of the Oppenheimer-Volkoff (OV) mass $M_{\rm OV}=0.384\, m_{\rm P} [m/m_{\rm P}]^{-2} = 0.816\, M$, OV radius $R_{\rm OV} = 3.35\, l_{\rm P} [m/m_{\rm P}]^{-2} = 3.56\, R$, and OV density $\rho_{\rm OV} = \SI{0.0196}{\rho_{\rm P}} [m/m_{\rm P}]^{4}= \SI{0.436}{\rho}$  \citep{2020EPJB...93..208A}.

Now we continue with the description of the fermionic mass distribution. The spherical symmetry is described by a metric $g_{\mu\nu}$ given in the standard form \begin{equation}
    g_{\mu\nu}(r) = \diag\parantheses{\e^{2\nu(r)}, - \e^{2\lambda(r)}, - \frac{r^2}{R^2} , - \frac{r^2}{R^2} \sin^2 \vartheta}.
\end{equation} The fermionic gas is assumed to be a perfect fluid described by the stress tensor \begin{equation}
    T^{\mu\nu}(r) = \diag\parantheses{\rho(r) c^2 , P(r), P(r), P(r)}.
\end{equation} The components of the stress tensor $T^{\mu\nu}(r)$ contain the mass density $\rho(r)$ and the isotropic pressure $P(r)$.

The metric $g_{\mu\nu}(r)$ and the stress tensor $T^{\mu\nu}(r)$ are sufficient to solve the Einstein equation. We obtain \begin{equation}
    \label{eqn:lambda:solution}
    \e^{-2 \lambda(r)} = 1 - \frac{R}{r}\frac{M(r)}{M}
\end{equation} and two equations for the mass $M(r)$ and metric potential $\nu(r)$ describing the distribution of self-gravitating fermions, \begin{equation}
    \label{eqn:mass}
    \diff{}{r/R} \frac{M(r)}{M} = \frac{\rho(r)}{\rho} \frac{r^2}{R^2}
\end{equation} and \begin{equation}
    \label{eqn:metric-potential}
    \diff{\nu}{r/R} = \frac12 \frac{R^2}{r^2}\brackets{\frac{M(r)}{M} + \frac{r^3}{R^3} \frac{P(r)}{\rho c^2}}\brackets{1 - \frac{R}{r} \frac{M(r)}{M}}^{-1}.
\end{equation} These are the Tolman-Oppenheimer-Volkov (TOV) equations \citep{1983bhwd.book.....S}. They express the condition of hydrostatic equilibrium $\d P = -\frac12(\epsilon+P)\d\nu$ in general relativity.

The particle number density $n(r)$, mass density $\rho(r)$ and pressure $P(r)$ are given by \begin{align}
    \label{def:particle-density}
    \frac{n(r)}{n}  &= \frac{4}{\sqrt{\pi}} \int \diff{\epsilon}{p} f(r, \epsilon) \epsilon^2 \d\epsilon, \\
    \label{def:mass-density}
    \frac{\rho(r)}{\rho}  &= \frac{4}{\sqrt{\pi}} \int \diff{\epsilon}{p} f(r, \epsilon) \epsilon^3 \d\epsilon, \\
    \label{def:pressure}
    \frac{P(r)}{\rho c^2} &= \frac{4}{3\sqrt{\pi}} \int \brackets{\diff{\epsilon}{p}}^3 f(r, \epsilon) \epsilon^3 \d\epsilon,
\end{align} where $n = \rho/m$ is a scale factor for the particle number density, $\epsilon = \sqrt{1 + p^2}$ is the particle energy in units of $m c^2$, $p$ is the momentum in units of $m c$, $v=\tdiff{\epsilon}{p} = \sqrt{\epsilon^2 - 1}/\epsilon$ is the particle velocity and $f(r,\epsilon)$ is a distribution function describing the phase space density of a particle at position $r$ and energy $\epsilon$.

Further, we calculate the particle number $N(r)$ with \begin{equation}
    \label{n77}
    \frac{{N(r)}}{N} = \int_0^{r} \frac{n(r)}{n} \brackets{1 - \frac{R}{r}\frac{M(r)}{M}}^{-1/2} \frac{r^2}{R^2} \frac{\d{r}}{R}
\end{equation} where $N = M/m$ is a scale factor for the particle number. From \cref{eqn:mass} we obtain the enclosed mass $M(r)$ with \begin{equation}
    \label{n53}
    \frac{{M(r)}}{M} = \int_0^{r} \frac{\rho(r)}{\rho} \frac{r^2}{R^2} \frac{\d{r}}{R}.
\end{equation}

From the kinetic theory \cite{1998MNRAS.300..981C} (see also \cref{sec_grf} for details) it is possible to justify a truncated Fermi-Dirac DF of the form  \begin{equation}
    \label{rfking}
    f(r,\epsilon) = \frac{1 - \e^{\brackets{\epsilon - \varepsilon(r)}/\beta(r)}}{1 + \e^{\brackets{\epsilon - \alpha(r)}/\beta(r)}}
\end{equation} for $\epsilon \leq \varepsilon(r)$ and otherwise $f(r, \epsilon) = 0$. Particles with an energy larger than the escape energy $\varepsilon(r)$ are considered as lost and produce a cutoff in the DF. This is called the general relativistic fermionic King model.

This DF contains various information encoded in the temperature variable $\beta(r)$, the chemical potential $\alpha(r)$ and the cutoff energy $\varepsilon(r)$. Note that $\alpha(r)$ and $\varepsilon(r)$ include the rest-mass, compare also with \cref{eqn:chemical-potential:def,eqn:escape-energy:def} below. All three parameters are related to the metric potential $\nu(r)$ through the Tolman relation \citep{1930PhRv...36.1791T}, the Klein relation \citep{1949RvMP...21..531K} and the conservation of energy \citep{1989A&A...221....4M}, \begin{equation}
    \label{eqn:red-shift-relations}
    \diff{\ln \beta}{r/R} = \diff{\ln \alpha}{r/R} = \diff{\ln \varepsilon}{r/R} = - \diff{\nu}{r/R}.
\end{equation}

Further, following the work of \cite{2018PDU....21...82A}, we introduce the degeneracy parameter $\theta(r)$ and the cutoff parameter $W(r)$, both defined by \begin{align}
    \label{eqn:chemical-potential:def}
    \alpha(r)      & = 1 + \beta(r) \theta(r), \\
    \label{eqn:escape-energy:def}
    \varepsilon(r) & = 1 + \beta(r) W(r).
\end{align}

The escape energy $\varepsilon(r)$ has a monotonically decreasing behavior, see \cref{eqn:red-shift-relations}, and we may define a boundary radius $r_b$ such that $\varepsilon(r_b) = 1$ or, equivalently, $W(r_b) = 0$. With this definition we obtain $f(r \geq r_b, \epsilon) = 0$ for any particle energy $\epsilon$.

Assuming regular conditions at the center of a fermionic mass distribution, i.e. $M(0) = 0$, each solution is characterized by the initial condition $\beta_0 = \beta(0)$, $\theta_0 = \theta(0)$ and $W_0 = W(0)$ for a given particle mass $m$.

The central value $\nu_0=\nu(0)$ may be calculated for every solution by imposing the Schwarzchild metric at the boundary where $P(r_b) = \rho(r_b) = 0$. The pressure and density drop to zero at the boundary radius $r_b$ (and beyond) because the DF from \cref{rfking} drops to zero due to the collaps of the potential well, i.e. $\varepsilon(r \geq r_b) = 1$ such that in \cref{def:particle-density,def:mass-density,def:pressure} no particle energies contribute to the integration. The matching of the metric potential at the boundary can be done after the calculation of a mass distribution because \cref{eqn:mass,eqn:metric-potential} do not depend on $\nu(r)$. However, for this work $\nu_0$ is not required.

In summary, every fermionic mass distribution following \cref{eqn:mass,eqn:metric-potential,rfking,eqn:red-shift-relations} is described by four parameters. But only three  parameters ($\beta_0$, $\theta_0$, $W_0$) characterize the normalized mass distribution since the particle mass $m$ acts as a scaling parameter as encapsulated in \cref{eqn:scale:mass,eqn:scale:length,eqn:scale:density}.

We can obtain  the equations of the problem summarized in this section (distribution function, equation of state, TOV equations, Tolman-Klein relations) all at once from a MEP as detailed in \cite{2020EPJP..135..290C} and \cref{sec_grf}. They result from the cancellation of the first order variations of entropy ($\delta S=0$ for entropy extrema) at fixed particle number and mass-energy. This variational approach also provides us with a criterion of dynamical and thermodynamical stability determined by the sign of the second variations of entropy ($\delta^2S<0$ for entropy maximum). This is the condition of stability that we consider below.


\section{Method}
\label{sec:method}

We perform a stability analysis for the MW, following the original ideas presented in \cite{2015PhRvD..92l3527C} within Newtonian gravity, though this time we treat the thermodynamic stability in full general relativity, as addressed in \cite{2020EPJB...93..208A,2021MNRAS.502.4227A}.

In general, for a given particle mass $m$ we find a fermionic DM solution developing a quantum core, a plateau, and a halo --- each associated with some astrophysical constraint. For that solution we then extract the values for the DM halo particle number $N_{\rm DM}$ and a characteristic parameter $\mu$ (see below) necessary to compute the caloric curves (see \cref{sec:stability-instruction} for details). Along the caloric curve the quantum core mass is not constrained anymore such that for every solution we obtain different intervals of stable and unstable core masses. We then determine whether the mass of the central dark compact core of the MW (identified with SgrA*) falls on the stable or unstable branch depending on the chosen constraints for a given particle mass $m$.

In summary, our method consist of two steps: (1) find the solution for a given set of constraints as detailed in \cref{sec:choice-constraints} and (2) check if that solution is stable as instructed in \cref{sec:stability-instruction}.

\subsection{Choice of constraints}
\label{sec:choice-constraints}
For a given particle mass $m$ each fermionic DM solution is described by three constraints. For instance, in the work of \cite{2018PDU....21...82A} the MW has been characterized by a core mass $M_c$ and two constraints of the embedded mass at $\SI{12}{\kilo\parsec}$ and $\SI{40}{\kilo\parsec}$.

In this work, we consider a different set of constraints composed of the total DM halo mass $\SYMmassDM$ characterizing the DM halo, the Keplerian mass $M_K$ characterizing the embedded quantum core, and the plateau density $\rho_p$ characterizing the inner halo.

The DM halo mass $\SYMmassDM = M(r_b)$ is defined as the mass enclosed within the boundary radius $r_b$ where the density falls to zero, i.e. $\rho(r_b) = 0$. 

The plateau density is given by $\rho_p = \rho(r_p)$ with $r_p$ being the plateau radius, which is defined at the local minimum in the rotation curve $v_{\rm DM}(r)$.

The Keplerian mass is given by $M_K = M(r_p)$ and reflects the mass plateau in the $M(r)$ profile precisely where the degenerate quantum core has fully transitioned towards the classical regime. Similar to $M_c$ from previous works, it is a measure of the mass of the dark compact object SgrA* (usually assumed to be a SMBH) that resides at the galactic center. The Keplerian mass is always slightly larger ($\sim \SI{10}{\percent}$) than the core mass $M_c$. See \cref{sec:core-mass-extrema} for further details about $M_K$ and its bounds.

With these three constraints ($M_K$, $\SYMmassDM$, $\rho_p$) we obtain a one-dimensional family of solutions. In the astrophysically relevant region of such core-halo DM solutions the particle mass $m$ acts as a free parameter. Moreover, the larger the particle mass, the more compact the dense quantum core at given $M_K$, until it reaches the critical value for gravitational collapse into a BH (see e.g. \cite{2018PDU....21...82A}).

\subsection{Determining stability}
\label{sec:stability-instruction}

For every fermionic DM solution, described by the parameters $m$, $\rho_p$, $M_K$ and $\SYMmassDM$, we extract the total particle number $N_{\rm DM} = N(r_b)$ and the value of the thermodynamic parameter $\mu = \e^{W_0 - \theta_0}$ necessary to do the stability analysis \citep{2015PhRvD..92l3527C,2021MNRAS.502.4227A}. That is, with the same particle mass $m$ we compute the caloric curve for the extracted parameters $N_{\rm DM}$ and $\mu$ and check if the considered DM solution falls on the stable branch. It is important to mention that for the second step we apply a different set of constraints compared to the first step. That is, we switch from ($m$, $\rho_p$, $M_K$ and $\SYMmassDM$) to ($m$, $N_{\rm DM}$ and $\mu$). That latter set has one parameter less such that we obtain a monoparametric curve --- the caloric curve --- for every DM solution as described by four parameters.  

The caloric curve is usually constructed by plotting the inverse of the temperature ($1/\beta_b$) --- as observed at the boundary $r_b$ --- against the negative of the total energy or its normalization, i.e. the negative of the DM halo mass ($-\SYMmassDM$). For a better comparison between different caloric curves we consider the expression $1 - \SYMmassDM/(m N_{\rm DM})$ for the mass. That expression can be interpreted as a normalized binding energy which reduces to the usual energy in the non-relativistic limit. It describes a simple affine transformation of the mass without changing the topology of the caloric curve. 

Having a caloric curve, we apply the Poincaré-Katz turning point criterion to identify the points of thermodynamic stability change and, therefore, to determine the stable and unstable regions \citep{1978MNRAS.183..765K,1979MNRAS.189..817K} (see also Appendix C in \citep{2020EPJB...93..208A} and Appendix A in \citep{2021MNRAS.502.4227A}). This is a powerful and rigorous method which relies only on the derivatives of specific caloric curves (e.g. inverse temperature vs. total energy), without the need to explicitly calculate the (rather involved) second order variations of entropy.

The Poincaré-Katz turning point criterion can be reduced to the following \enquote{rule of thumb}: (a) In the microcanonical ensemble, a change of stability can occur only at a turning point of mass-energy. (b) One mode of stability is lost when the caloric curve rotates clockwise and gained when it rotates anticlockwise. Starting with a stable mode we obtain an unstable branch when the first extremum of the mass is reached. It remains unstable while the caloric curve rotates clockwise. (c) When the same curve starts to rotate anticlockwise then a stable mode is re-gained on every extremum of the mass. In this sense, once in a given unstable branch of the caloric curve (coming from an originally stable branch), it is necessary to have as many anticlockwise turns of the curve as clockwise passed to regain the thermodynamic stability.

\subsection{Choice of parameter values}
\label{sec:choice-param}

Of interest are the astrophysically relevant values for the parameters describing a DM solution. From previous works and observations we can use safe bounds as described in the following.

In \cite{2018PDU....21...82A} the particle mass bounds have been estimated to be in the range $mc^2 \approx \SIrange{48}{345}{\kilo\eV}$ using data of the Grand rotation curve of the Galaxy from \cite{2013PASJ...65..118S} plus including  the central S-cluster stars  \citep{2009ApJ...707L.114G} in the analysis. Later, in \cite{2020A&A...641A..34B,2021MNRAS.505L..64B}, a more sophisticated geodesic analysis of the S-cluster star was performed raising the lower bound to about $\SI{56}{\kilo\eV}$.

The lower bound can be understood qualitatively by considering the non-relativistic mass-radius relation $M=91.9\, \hbar^6/G^3m^8R^3$ of a degenerate fermion ball with mass $M=\SI{4.2E6}{\Msun}$ and requiring that $R<\SI{6E-4}{\parsec}$, the S2 star pericenter, yielding $m > \SI{54.6}{\kilo\eV}$.

The upper bound is reached when the mass of the quantum core (fermion ball) reaches the critical OV mass $M_{\rm OV}=0.384\, m_{\rm P}^3/m^{2}$ and collapses towards a BH. An upper bound of about $\SI{345}{\kilo\eV}$ is obtained when the dark compact object in the Galactic center is described with the quantum core mass $M_c$ \citep{2018PDU....21...82A}. When the compact object is instead described with the Keplerian mass $M_K$ --- as is done in this work ---  the upper limit increases to about $\SI{387}{\kilo\eV}$.

Therefore, for our stability analysis we consider three values for the particle mass: the lower bound $\SI{56}{\kilo\eV}$, the upper bound $\SI{387}{\kilo\eV}$, and additionally a particle mass of $mc^2 = \SI{250}{\kilo\eV}$ as a convenient value in between.

For the core mass, associated with the Keplerian mass $M_K$, we proceed similarly to \cite{2018PDU....21...82A} and consider a fixed value ${M_K}^\text{\tiny (MW)} = \SI{4.2E6}{\Msun}$ since it is well constrained by observations \citep{2009ApJ...707L.114G,2019A&A...625L..10G}.

The total mass $M_{\rm tot}$ of the MW (including baryons) is less constrained but is estimated to be of the order of $\SIrange{2E11}{5E11}{\Msun}$ \citep{2023A&A...678A.208J}. Nevertheless, for the DM halo mass $\SYMmassDM$ as being the dominant component we consider values from a wider range in the interval $\SIrange{5E10}{1E13}{\Msun}$ with the purpose to get a broader picture about the stability behavior.

Similarly interesting but poorly constrained from observations is the plateau density $\rho_p$. From the work of \cite{2018PDU....21...82A} we choose a DM plateau density $\rho_p = \SI{1E-2}{\Msun \per \parsec\cubed}$ as a convenient order-of-magnitude representative of the MW for the stability analysis. In \cref{sec:results:comparison} we consider a different $\rho_p = \SI{2E-2}{\Msun \per \parsec\cubed}$ adapted to recent estimations from GAIA DR3 \cite{2021A&A...653A..86W}.

We remind that those bounds (i.e. $\rho_p$) are considered only for the first methodological step of finding a DM distribution resembling the observed Galactic parameters. In the second step those constraints do no apply anymore such that along the caloric curve there is only one solution which fulfills the constraints of the first step.


\section{Results}
\label{sec:results}

In this section we determine the stability behavior of different core-halo solutions, with the aim to constrain the DM particle mass range for specific DM configurations which agree with the MW observables.

\subsection{Stability analysis}
\label{sec:results:stability-analysis}

\begin{figure*}%
    \centering%
    \includegraphics[width=\hsize]{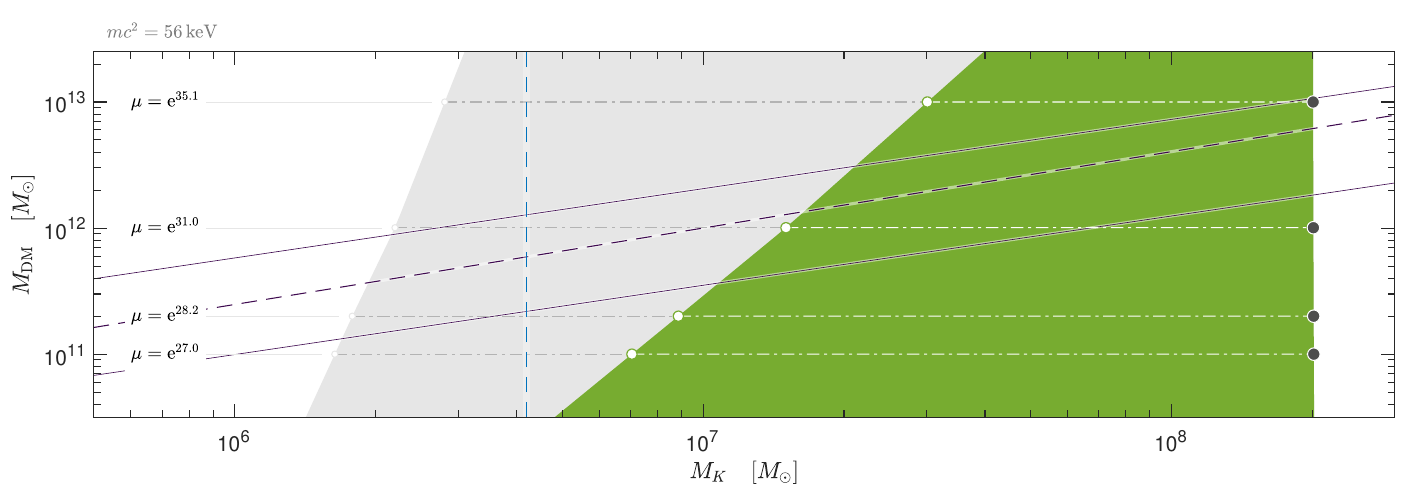}
    \includegraphics[width=\hsize]{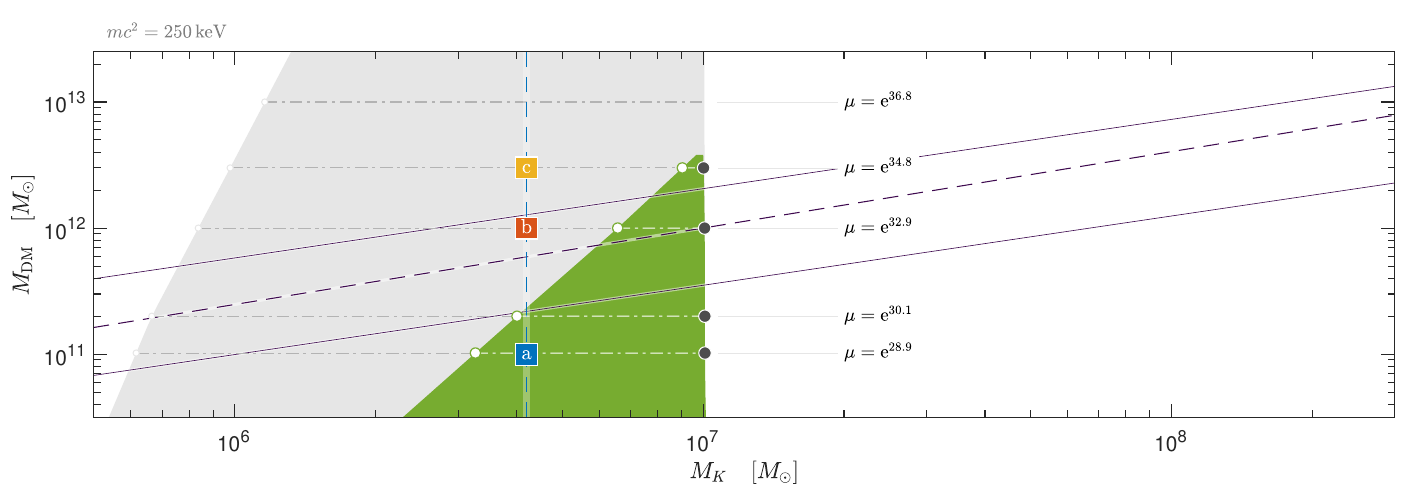}
    \includegraphics[width=\hsize]{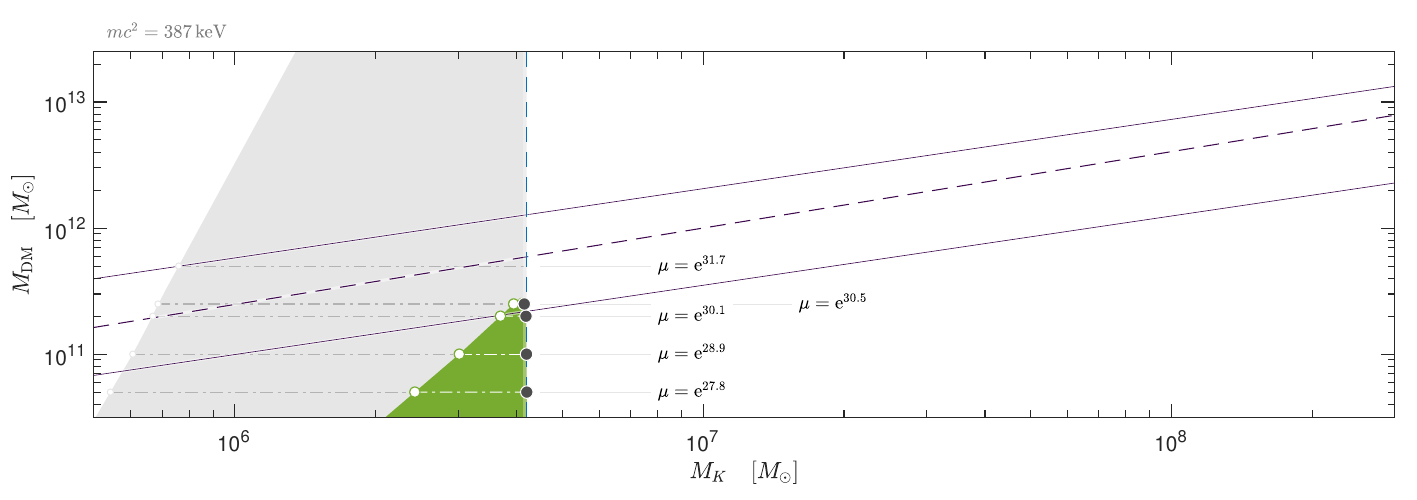}
	\caption{
        Stability analysis for a particle mass of \SI{56}{\kilo\eV}, \SI{250}{\kilo\eV} and \SI{387}{\kilo\eV}, each with specific constraints $N_{\rm DM}$ and $\mu=\e^{W_0 - \theta_0}$. See \cref{sec:method} how the constraints are obtained from MW observations and how caloric curves are computed. From the caloric curves we extract the Keplerian (core) mass $M_K$ and the DM halo mass $M_{\rm DM}$. Stable core-halo solutions are represented by the green area while unstable core-halo solutions are represented by the gray area. Dot-dashed lines show the computed values. All other values are interpolated. The vertical dashed line represents a MW-like galaxy with a core mass ${M_K}^\text{\tiny (MW)} = \SI{4.2E6}{\Msun}$. All plots show the same ($M_K$, $\SYMmassDM$)-window for better comparison how the areas change for different particle masses. The green triangle (stable region) shifts to the left as $m$ increases until it saturates to ${M_K}^\text{\tiny (max)}={M_K}^\text{\tiny (MW)}$ at the maximum particle mass $mc^2=\SI{387}{\kilo\eV}$. The core-halo relation found by \citet{2002ApJ...578...90F} is shown as purple dashed line together with the bounds as purple solid lines.
    }%
	\label{fig:mkms:rhop-10}%
\end{figure*}

\begin{figure*}%
	\centering%
	\includegraphics[width=\hsize]{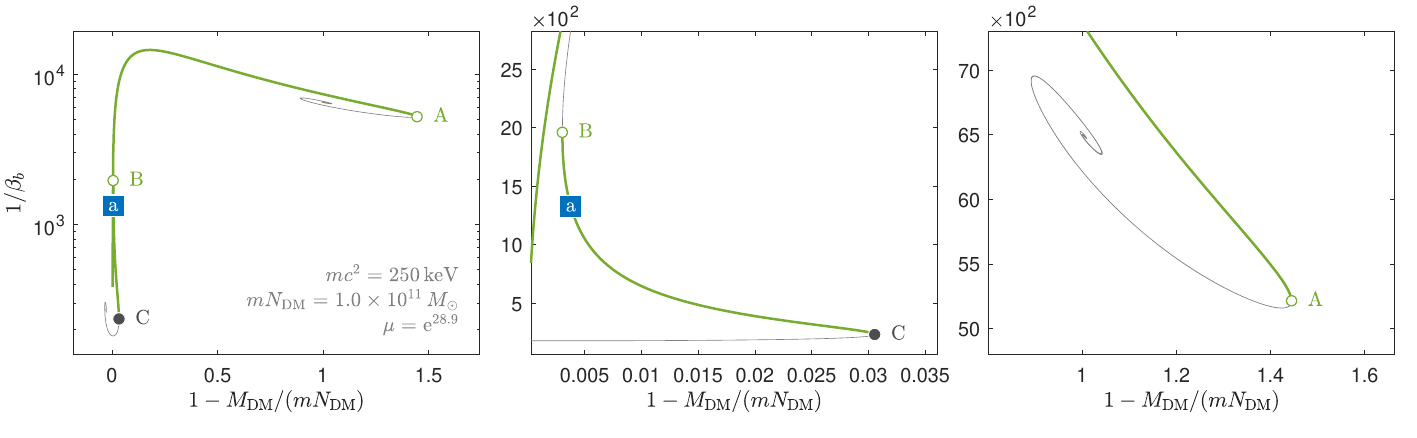}
	\includegraphics[width=\hsize]{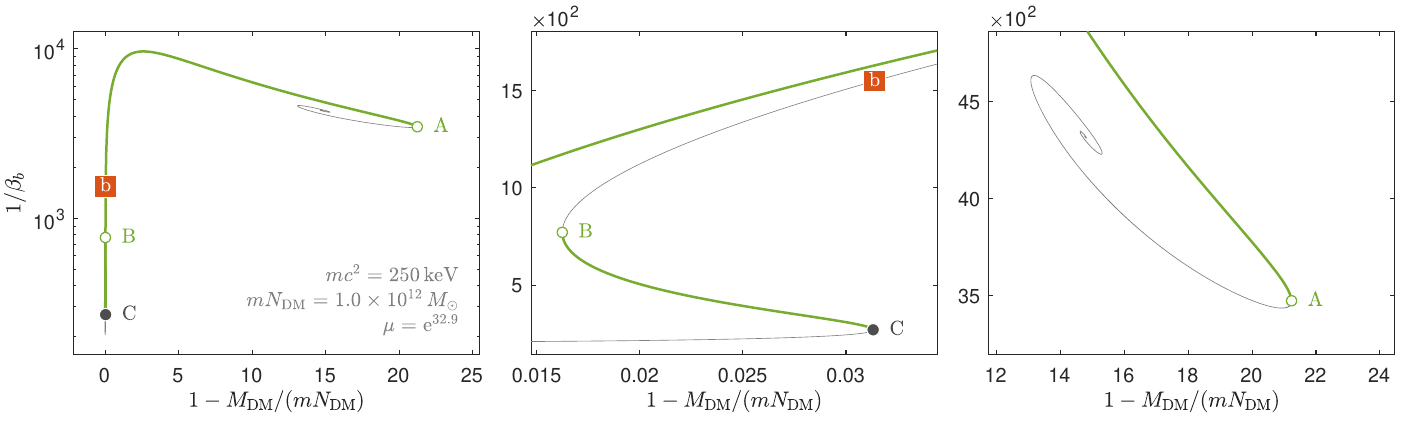}
	\includegraphics[width=\hsize]{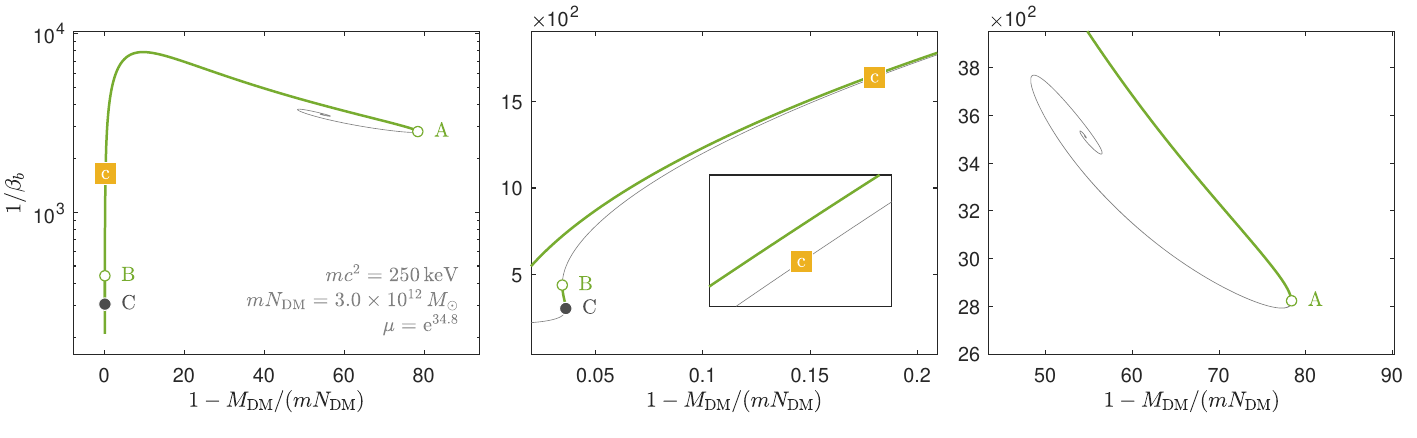}
	\caption{Caloric curves corresponding to the three examples (a, b, c) with fixed DM halo masses in \cref{fig:mkms:rhop-10} for the case of \SI{250}{\kilo\eV}. Thin lines represent unstable solutions. Thick lines represents stable solutions. Points A, B and C indicate a change of stability. The additional window in the bottom panel clarifies that solution (c) belongs to the unstable branch. The extension of the stable core-halo branch (between B and C) becomes smaller and smaller from top to bottom panel.}%
	\label{fig:caloric-curves:250}%
\end{figure*}

Let us first discuss the case of a DM particle mass of $mc^2 = \SI{250}{\kilo\eV}$ in full detail to describe the general stability behavior. The corresponding stability diagram in the $(M_K,\SYMmassDM)$ plane is shown in the second panel of \cref{fig:mkms:rhop-10}. Additional illustrative caloric curves are shown in \cref{fig:caloric-curves:250} for different values of $\SYMmassDM$.

The stability diagram is a convenient representation which combines different caloric curves containing the information about stability. The caloric curves in \cref{fig:caloric-curves:250} display points of stability change (corresponding to turning points of mass-energy) which we label as $A$, $B$ and $C$. For core-halo solutions only points $B$ and $C$ are relevant. Point $A$ corresponds to diluted halo-only solutions without a degenerate nucleus (see \cref{sec:method}, see also Appendix A in \cite{2021MNRAS.502.4227A} for further details on stability change).

In \cref{fig:mkms:rhop-10} we focus on a window for $\SYMmassDM$ to the relevant range of about $\SIrange{5E10}{2E13}{\Msun}$. The computed values are shown as horizontal dot-dashed lines, i.e. fixed $\SYMmassDM$. All values in between on the vertical axis are interpolated and extrapolated slightly beyond the considered range. This way, we construct a stability diagram where the thermodynamically stable solutions are represented by a green area (having the shape of a triangle) while the  thermodynamically unstable solutions are represented by a grey area. The white areas indicate the absence of core-halo solutions for the chosen set of constraints.

For a fixed DM halo mass $\SYMmassDM$ the domain of stability (green area) goes from point B (open circle) to point C (black-filled circle). The core-halo solutions with $M_K \in [{M_K}^\text{\tiny (B)}, {M_K}^\text{\tiny (C)}]$ are thermodynamically stable, i.e. entropy maxima at fixed particle number and mass-energy. The core mass ${M_K}^\text{\tiny (C)}$ at point C corresponds to the OV limit $M_{\rm OV}$ of marginal stability (see \cref{sec:core-mass-extrema}). At that point the core mass becomes dynamically unstable in the sense of general relativity and is expected to collapse towards a SMBH. It represents the maximal stable core mass. No stable core-halo solutions exist for $M_K > {M_K}^\text{\tiny (C)}$. On the other hand, the core mass ${M_K}^\text{\tiny (B)}$ represents the minimal stable core mass. The core-halo solutions with $M_K < {M_K}^\text{\tiny (B)}$ are thermodynamically unstable, i.e. saddle points of entropy at fixed particle number and mass-energy.

The unstable domain (grey area) is limited by the minimum ${M_K}^\text{\tiny (0)}$, corresponding to a solution lying within the upper spiral of the caloric curve, just when the spiral starts to unwind (see \cref{sec:core-mass-extrema}). The unwinding of the classical spiral when quantum effects are accounted for was first realized in \cite{1998MNRAS.296..569C} within Newtonian gravity, and further corroborated within general relativity in \cite{2020EPJB...93..208A,2021MNRAS.502.4227A}, leading to the avoidance of the gravothermal catastrophe of collisionless nature.

Below the threshold ${M_K}^\text{\tiny (0)}$, the concept of a quantum core mass does not apply any longer. That is, mildly degenerate core masses below ${M_K}^\text{\tiny (0)}$ correspond to low central degeneracy parameters ($\theta_0 \lesssim 7$), further implying that the thermal de Broglie wavelength is smaller than twice the inter-particle mean distance at such mildly degenerate cores. Thus, the quantum statistical condition for the core no longer applies and the fermionic solution transitions towards the diluted (Boltzmannian) regime where $\theta_0 \ll -1$ (see \cite{2021MNRAS.502.4227A} for further details).

The stability region $[{M_K}^\text{\tiny (B)}, {M_K}^\text{\tiny (C)}]$ shrinks as the mass $\SYMmassDM$ of the DM halo increases. Above a maximum ${\SYMmassDM}^\text{\tiny (max)}$ there are no stable core-halo solutions whatever the value of $M_K$. This corresponds to the case where the turning points B and C of the caloric curve have merged so there is no condensed branch anymore (see \cref{fig:caloric-curves:250}).

Let us now consider briefly the stability of the other DM particle masses by focusing on the stability diagram in \cref{fig:mkms:rhop-10}.

For a particle mass of $mc^2 = \SI{56}{\kilo\eV}$ \citep{2020A&A...641A..34B} we see that the vertical dashed line (MW line) is on the outside left of the green area. To be precise, stability is reached only for very low DM halo masses ($\lesssim \SI{E10}{\Msun}$). Therefore, the MW core-halo solutions are never stable in the relevant range ($\sim\SI{E11}{\Msun}$) of the DM halo mass $\SYMmassDM$ \citep{2023A&A...678A.208J}. This result suggests that a too light fermionic particle mass should be rejected as a DM candidate. That is, for $mc^2 = \SI{56}{\kilo\eV}$ it does not exist any stable fermionic solution able to account for the compact object at the center of the MW together with an astrophysically relevant DM halo.

As the particle mass $m$ increases the green area moves effectively to the left while shrinking also in size. For some particle mass being large enough the intersection of MW core solutions (e.g. the vertical dashed line in \cref{fig:mkms:rhop-10}) and the MW halo solutions (i.e $\SYMmassDM \sim \SI{E11}{\Msun}$) fall in the green area. For those particle masses the MW core-halo solutions are stable as discussed above in the specific case of $mc^2=\SI{250}{\kilo\eV}$.

For the maximal particle mass of $mc^2 = \SI{387}{\kilo\eV}$ the MW core mass ${M_K}^\text{\tiny (MW)} = \SI{4.2E6}{\Msun}$ coincides with the maximal core mass (point C). As a result, the vertical dashed line (MW line in \cref{fig:mkms:rhop-10}) is at the extreme right of the green area. Therefore, the MW core-halo solutions are always marginally stable for DM halo masses $\SYMmassDM < {\SYMmassDM}^\text{\tiny (max)}$.

For $mc^2 > \SI{387}{\kilo\eV}$ the quantum core is gravitationally unstable with respect to general relativity and collapses. This result suggests that if the DM particle is too heavy then the MW should harbor a BH rather than a fermion ball.

We compare our results also with the ``fundamental relation between SMBH and DM halos'' found by \cite{2002ApJ...578...90F}. According to this study, the mass of the BH (which we identify as our fermionic nucleus) and the mass of the hosting halo lie on a narrow stripe following a relation given by \begin{equation}
    \frac{M_{\rm BH}}{\SI{E8}{\Msun}} \approx 0.1 \brackets{\frac{\SYMmassDM}{\SI{E12}{\Msun}}}^{1.65}.
\end{equation} A similar relation has been obtained in the work of \cite{2015ApJ...800..124B}. It is remarkable that the stable region (green area in \cref{fig:mkms:rhop-10}) from our stability analysis has always an intersection with the narrow stripe. Therefore, we can always find solutions which are in agreement with the $M_{\rm BH}$-$\SYMmassDM$ relation in a given halo mass range. For DM core masses above the stability region, the central object has become a BH which can subsequently grow towards larger masses \citep{2023MNRAS.523.2209A}.

\subsection{Morphological features of DM halos}
\label{sec:result:morphological-features}

In the following, we take a closer look at three fermionic benchmark solutions from \cref{sec:results:stability-analysis} with different halo masses for the case of \SI{250}{\kilo\eV}, i.e. along the vertical dashed-line in the middle panel of \cref{fig:mkms:rhop-10}. For these three examples --- labeled as $(a)$, $(b)$, $(c)$ --- we show in \cref{fig:caloric-curves:250} the corresponding caloric curves.

For a DM halo mass of ${\SYMmassDM}^\text{\tiny (a)} = \SI{1E11}{\Msun}$ solution $(a)$ is within the stable branch. For ${\SYMmassDM}^\text{\tiny (b)} = \SI{1E12}{\Msun}$ solution $(b)$ is within the unstable branch, though there exist stable solutions with larger core masses (compare with \cref{fig:mkms:rhop-10}). For ${\SYMmassDM}^\text{\tiny (c)} = \SI{3E12}{\Msun}$ the halo mass is just below the maximal halo mass ${\SYMmassDM}^\text{\tiny (max)}$ such that there is only a very narrow range for the core mass where the solutions are stable --- see intersection of dot-dashed horizontal line having $\mu=e^{34.8}$ with the tip region of the green triangle in \cref{fig:mkms:rhop-10}. Nevertheless, for the shown solution $(c)$ the core mass is considerably lower than the narrow range falling in the desired stable region. For larger halo masses above the maximal mass threshold the entire mass distribution becomes too massive such that no stable branches exist. In that case, we expect the formation of a SMBH instead of a fermion-core.

\begin{figure}%
	\centering%
	\includegraphics[width=\hsize]{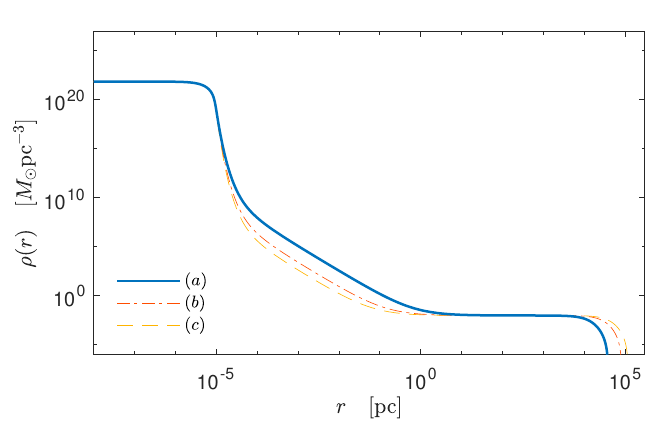}
	\caption{Illustration of three density profiles for the case of \SI{250}{\kilo\eV}, see colored squares in \cref{fig:mkms:rhop-10}. (a) represents a stable solution (thick solid line), (b) is in the unstable region (thin dot-dashed line), (c) is in the unstable region with $\SYMmassDM \lesssim {\SYMmassDM}^\text{\tiny (max)}$ (thin dashed line).}%
	\label{fig:density:250}%
\end{figure}

In \cref{fig:density:250} we show the corresponding density profiles for the three examples $(a)$, $(b)$, and $(c)$. All solutions have the same nucleus (as constrained by $M_K$) followed by the same plateau (as constrained by $\rho_p$). A similar halo in all three examples is described by a polytrope with index $\nicefrac{5}{2}$. Hence, the differences are the transition from the nucleus to the plateau (over the range $\SIrange[print-zero-exponent]{E-5}{E0}{\parsec}$) and the size of the halo ($\sim\SI{E5}{\parsec}$). From the stability analysis we find only solution $(a)$ to be stable.

A typical interest is the morphology of the outer halo, in particular the slope of the outer density profile, and whether any characteristics may indicate a stable behavior. At this point we need to provide a different picture because the stability change occurs far in the polytropic regime ($W_p \ll 1$) where the outer halo has practically a unique shape, see e.g. \cref{fig:density:250}. Therefore, we focus on the plateau cutoff parameter $W_p$ which is a good representative of the characteristics of the halo. 

\begin{figure}%
	\centering%
	\includegraphics[width=\hsize]{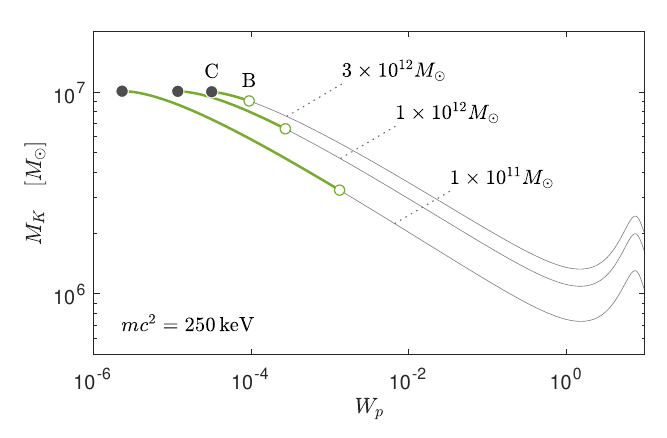}
	\caption{$W_p$-$M_K$ curve corresponding to selected horizontal lines of the second panel in \cref{fig:mkms:rhop-10}. Shown are the stable branches bounded by the points B (green circle) and C (black dot) and the unstable branches before point B. The value of $W_p$ allows us to describe the characteristics of the outer halo and it is sufficient to show only up to $W_p = 10$. For clarity, we do not show the unstable branch after point C.}%
	\label{fig:Wp:250}%
\end{figure}

Still considering the case of a particle mass of $\SI{250}{\kilo\eV}$, we plot in \cref{fig:Wp:250} the values of cutoff parameter $W_p$ and core mass $M_K$, both taken along the caloric curve for a given DM halo mass $\SYMmassDM$. That is the values correspond to the horizontal lines of the second panel in \cref{fig:mkms:rhop-10}. From the stable branches (thick green lines) we extract the lower and upper bounds of $W_p$ and $M_K$. See \cref{tbl:Wp} for numerical values.

With the information of $W_p$ we can tell whether the outer halo is polytropic with index $\nicefrac{5}{2}$ (for $W_p \ll 1$), or isothermal-like (for $W_p \gtrsim 10$). Interestingly, this fermionic theory for DM halos also allows for power law-like halo tails including $\rho \sim r^{-3}$ for $W_p \approx 7$. However they are unstable for core-halo morphologies (as shown in \cref{fig:Wp:250}) while they can be stable only in the diluted regime, the latter not analyzed in this work. See \cite{2023ApJ...945....1K} for further discussion and Fig. 6 in that paper for further details about the morphology of the fermionic halos.

From this morphology analysis, we find that the upper bounds for the three benchmark solutions are far in the polytropic regime with $W_p \lesssim \num{E-3}$. Hence, considering only the shape of the halo does not allow us to determine the stability behavior. Instead, the core must be taken into account and contrasted with the halo with associated $W_p$ values. Only when the halo reveals a flat tail (i.e. being isothermal) and harbors a quantum core then such a fermionic DM distribution is certainly unstable. This result is in line with the fact that isothermal-like (i.e. $\rho \sim r^{-2}$) halo tails are typically disfavored by observations. 

\begin{table}
    \caption{Upper and lower bounds of the core mass $M_K$ and the cutoff parameter $W_p$ for a given DM halo mass $\SYMmassDM$. The values have been numerically extracted from the stability analysis of solutions constrained by $mc^2 = \SI{250}{\kilo\eV}$, $M_K = \SI{4.2E6}{\Msun}$ and $\rho_p = \SI{E-2}{\Msun\per\parsec\cubed}$. All values are rounded. They correspond to the stable branches in \cref{fig:mkms:rhop-10} (second panel) and \cref{fig:Wp:250}.}
    \centering
    \begin{tblr}{
        width = \hsize,
        colspec = {
            *{1}{Q[1, si={table-format=1.0e1}]}
            *{3}{Q[1, si={table-format=1.1e1}]}
            *{2}{Q[1.1, si={table-format=1.1e1}]}
        }
    }
        \toprule
        ${\SYMmassDM}^\text{\tiny (B)}$ & ${M_K}^\text{\tiny (B)}$ & ${M_K}^\text{\tiny (C)}$ & ${W_p}^\text{\tiny (C)}$ & ${W_p}^\text{\tiny (B)}$\\
        $\textcolor{gray}{\si{\Msun}}$ & $\textcolor{gray}{\si{\Msun}}$ & $\textcolor{gray}{\si{\Msun}}$ & & \\
        \midrule
        1.0E11 & 3.3E6 & 1.0E7 & 2.3E-6 & 1.3E-3\\
        1.0E12 & 6.6E6 & 1.0E7 & 1.2E-5 & 2.7E-4\\
        3.0E12 & 9.0E6 & 1.0E7 & 3.2E-5 & 9.4E-5\\
        \bottomrule
    \end{tblr}
    \label{tbl:Wp}
\end{table}

\subsection{Comparison with recent MW observations}
\label{sec:results:comparison}

The aim of this section is to focus on more updated MW observables. That is, we will focus solely on thermodynamically stable solutions which include a dense fermionic nucleus of $\SI{4.2E6}{\Msun}$ mimicking the central dark compact object, together with constraints from most recent GAIA DR3 rotation curve data. 

In particular, we explain the total rotation curve as a superposition of baryonic components and a stable fermionic DM component. While the parameters of the DM component remain fixed, we adjust the baryonic mass model parameters following a best-fit approach to fit the MW rotation curve with data composed from different regions. For the halo (also partially covering the disk region) we take data from \cite{2023A&A...678A.208J}. For the intermediate region, covering bulges and disk, we take data from \cite{2013PASJ...65..118S}. For the central region we add observables from the best resolved S-cluster stars from \cite{2009ApJ...707L.114G} indicating the presence of the supermassive dark compact object.

Since the adopted galactocentric values $(R_\odot, V_\odot)$ in our considered rotation curves differ, i.e. $(\SI{8}{\kilo\parsec}, \SI{200}{\kilo\metre\per\second})^\text{\tiny Sofue}$ and $(\SI{8.34}{\kilo\parsec}, \SI{220}{\kilo\metre\per\second})^\text{\tiny Jiao}$, we corrected the observed rotation velocity data. Following the approach explained in \cite{2013PASJ...65..118S}, we apply the transformations \begin{align}
    r^c &= r \frac{{R_\odot}^\text{\tiny (Jiao)}}{{R_\odot}^\text{\tiny (Sofue)}},\\ 
    v^c(r) &= v(r) + \Delta v \frac{r}{{R_\odot}^\text{\tiny (Sofue)}}, 
\end{align} onto the data from \cite{2013PASJ...65..118S} to be properly comparable to the data from \cite{2023A&A...678A.208J}. $\Delta v = {V_\odot}^\text{\tiny (Jiao)} - {V_\odot}^\text{\tiny (Sofue)} = \SI{20}{\kilo\metre\per\second}$ is the difference of the solar velocities. For clarity, we consider data from \cite{2013PASJ...65..118S} only up the the first data point from \cite{2023A&A...678A.208J}.

For the DM component we consider --- as before --- a fermionic core with a mass of ${M_K}^\text{\tiny (MW)} = \SI{4.2E6}{\Msun}$ \citep{2009ApJ...707L.114G,2019A&A...625L..10G}. Motivated by the analysis of stellar streams in \cite{2024A&A...689A.194M} done within the fermionic DM framework we set the DM halo mass ${\SYMmassDM}^\text{\tiny (MW)} = \SI{1.4E11}{\Msun}$ being below the most likely total mass (including baryons) of about $\SI{2E11}{\Msun}$ as inferred from most recent GAIA DR3 rotation curve data \citep{2023A&A...678A.208J}. For the plateau density, we consider an adapted value of $\rho_p = \SI{2E-2}{\Msun\per\parsec\cubed}$ to be more in line with recent estimations for the local DM density \citep{2021A&A...653A..86W} and similar to earlier estimations \cite{2010A&A...523A..83S,2013JCAP...07..016N}. Further, we consider exemplary a DM particle mass of $mc^2 = \SI{250}{\kilo\eV}$. See \cref{sec:results:mass-implications} for other DM particle mass candidates.

\begin{figure}%
	\centering%
	\includegraphics[width=\hsize]{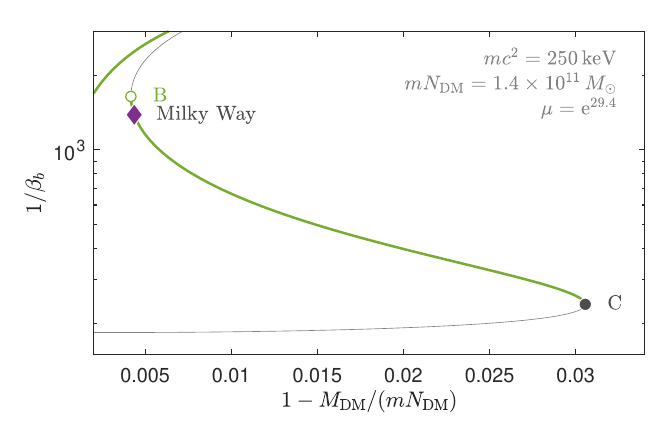}
	\caption{Caloric curve for a DM distribution described by the parameters $mc^2=\SI{250}{\kilo\eV}$, $M_K = \SI{4.2E6}{\Msun}$, $\SYMmassDM = \SI{1.4E11}{\Msun}$ and $\rho_p = \SI{2E-2}{\Msun\per\parsec\cubed}$. These parameters represent a stable DM configuration for the MW.}%
	\label{fig:caloric-curve:mw}%
\end{figure}

From the caloric curve shown in \cref{fig:caloric-curve:mw}, we demonstrate that such a DM distribution characterized by the chosen configuration parameters falls within the stable branch of the fermionic halos. Moreover, as explicitly shown for the Milky Way case in \cref{fig:caloric-curve:mw} (for $m=250$ keV; see purple diamond), the stable core-halo solution is quite far from the OV limit of gravitational collapse into a BH (point C), indeed is closer to point B, at the onset of stability change (see also \cite{2021MNRAS.502.4227A} for a discussion about the long-lifetime of such stable states).

\begin{figure*}%
	\centering%
	\includegraphics[width=\hsize]{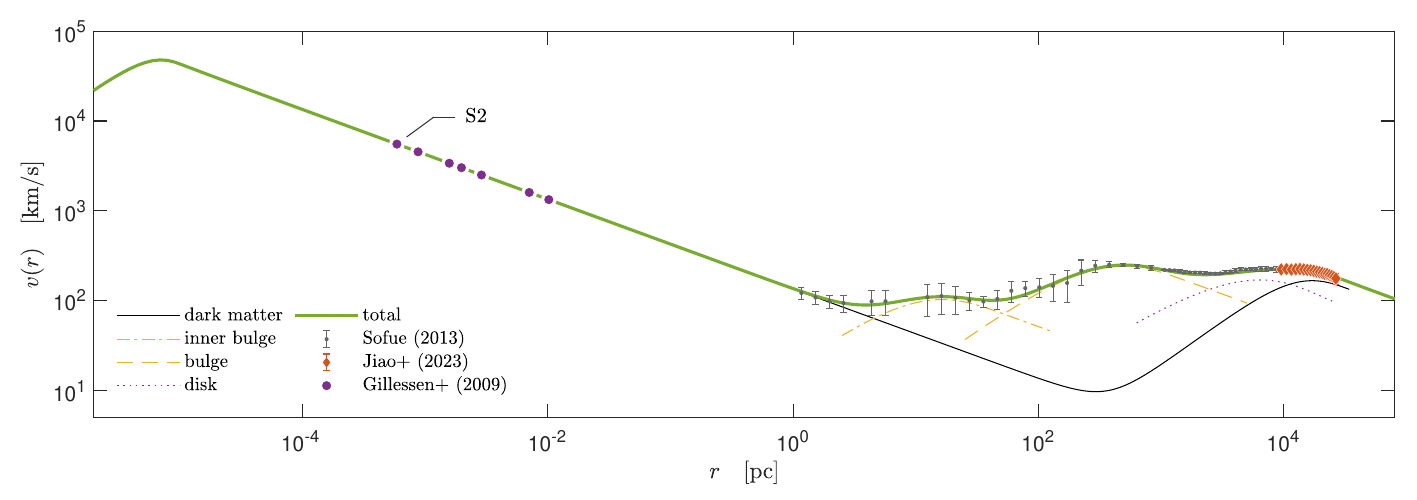}
	\caption{Illustration of the Galactic rotation curve and its baryonic components following \cite{2013PASJ...65..118S} with adapted values from a best-fit. For the DM component we consider the stable case for a particle of mass $\SI{250}{\kilo\eV}$ with $\SYMmassDM = \SI{1.4E11}{\Msun}$, $M_{K} = \SI{4.2E6}{\Msun}$ and $\rho_{p} = \SI{2E-2}{\Msun\per\parsec\cubed}$. For clarity, we do not show the outer regions with a power-law behavior, i.e. we highlight only the maximum of each baryonic component. On halo scales we add further observables from the latest Gaia-DR3 results \citep{2023A&A...678A.208J}. In the inner region we add some observables from the S-cluster \citep{2009ApJ...707L.114G} indicating the presence of a compact nucleus.}%
	\label{fig:rotation-curve:250}%
\end{figure*}

\begin{figure}%
	\centering%
	\includegraphics[width=\hsize]{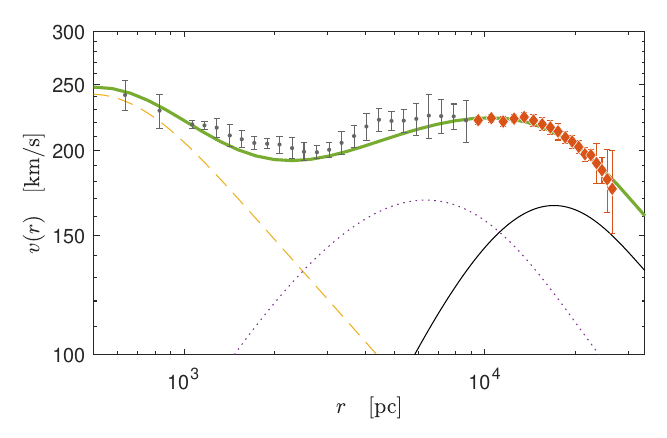}
	\caption{Zoom on the halo in \cref{fig:rotation-curve:250} showing a remarkable agreement with the GAIA DR3 rotation curve data.}%
	\label{fig:rotation-curve-zoom:250}%
\end{figure}

The rotation curve of a fermionic DM distribution is then given by \begin{equation}
    \brackets{\frac{v_{\rm DM}(r)}{c}}^2 = \frac{r}{R} \diff{\nu}{r/R}.
\end{equation}
Far from the dense core, the local relativistic velocity shown by this formula tends to the classical limit $v^2(r) = G M(r)/r$.

For the baryonic components (inner bulge, main bulge, disk), we follow the mass models in \cite{2013PASJ...65..118S}. The inner and main bulge components are described by an exponential sphere with a density profile given by $\rho_b(r) = \rho_b \e^{-r/R_b}$. $R_b$ and $\rho_b$ are the length and density scale factors, respectively, and are specific for each bulge component. The mass profile $M_b(r)$ is then given by $\d M_b = 4\pi r^2 \rho_b(r) \d r$ and the total bulge mass becomes $M_b = 8\pi R_b^3 \rho_b$. The disk is described by an exponential (Freeman) disk with a surface density given by $\Sigma_d(r) = \Sigma_d \e^{-r/R_d}$. $R_d$ and $\Sigma_d$ are the length and surface density scale factors, respectively. The mass profile $M_d(r)$ is then given by $\d M_d = 2\pi r \Sigma_d(r) \d r$ and the total disk mass becomes $M_d = 2\pi R_d^2 \Sigma_d$.

The rotation curve of each baryonic component $i$ follows from Newtonian dynamics, ${v_i}^2(r) = G M_i(r) / r$, and the grand (total) rotation curve is then simply a superposition of all components, \begin{equation}
    {v_{\rm tot}}^2(r) = {v_{\rm DM}}^2(r) + \sum_i {v_i}^2(r).
\end{equation}

In the grand rotation curve, see \cref{fig:rotation-curve:250}, we cover multiple orders of magnitude, from the galactic center to the outer halo. In \cref{fig:rotation-curve-zoom:250} we focus on the halo with recent results from GAIA DR3 studies. The best-fitted parameters for the baryonic components are listed in \cref{tbl:mw:parameters}.

The relatively large disk mass value is in line with inferences by \cite{2017A&A...598A..66P} and marginally in line with \cite{2023A&A...678A.208J}, while the DM halo density at the solar neighborhood ($\SI{8.1}{\kilo\parsec}$ from the center) is approximately $\SI{1E-2}{\Msun\per\parsec\cubed} \approx \SI{0.4}{\giga\eV \per \speedoflight\squared \per \centi\metre\cubed}$ which agrees (within $1\sigma$) with the value inferred in \cite{2021A&A...653A..86W}. This local DM density is comparable to but lower than the plateau density $\rho_p$ because the solar neighborhood falls into the halo region where the DM density starts to decrease.

Finally we comment on the (potential) effects due to the baryons in-fall onto the DM phase-space density. Such effects have been traditionally studied in the context of CDM halos, for example by introducing the concept of the Cole-Binney core \citep{2017MNRAS.465..798C} in realistic modeling of the MW mass components \citep{2023MNRAS.520.1832B}. However, as explained in \cite{2023ApJ...945....1K,2023Univ....9..197A}, our fermionic particles decouple from the primordial plasma while relativistic (i.e. 'a la WDM') and its relaxation towards equilibrium is achieved through a Maximum Entropy Principle (MEP), already leading to 'cored' inner halo profiles. Considering that it has been shown that baryonic feedback effects are considerably milder in cored WDM-like halos than for cuspier CDM halos (see e.g. \cite{2019MNRAS.483.4086B}), it remains to be proven in the context of our fermionic candidates (though out of the scope of this work) if the baryonic effects reinforce the existence of our core (or plateau).

\begin{table}
    \caption{Best-fit parameters for the baryonic components of the MW when superposed with a given fermionic DM component characterized by $\rho_p = \SI{2E-2}{\Msun\per\parsec\cubed}$, $M_K = \SI{4.2E6}{\Msun}$, $\SYMmassDM = \SI{1.4E11}{\Msun}$ and $mc^2 = \SI{250}{\kilo\eV}$.}
    \centering
    \begin{tblr}{
        width = \hsize,
        colspec = {
                 X[1, l]
            *{2}{Q[1, si={table-format=1.1e1}]}
        }
    }
        \toprule
         & \text{total mass} & \text{length scale} \\
         & $\textcolor{gray}{\si{\Msun}}$ & $\textcolor{gray}{\si{\parsec}}$ \\
        \midrule
        inner bulge & 5.9E7 & 4.7\\
        bulge & 1.0E10 & 1.4E2\\
        disk & 5.1E10 & 2.7E3\\
        \bottomrule
    \end{tblr}
    \label{tbl:mw:parameters}
\end{table}

\subsection{Implications on the DM particle mass}
\label{sec:results:mass-implications}

It is worth taking now a more detailed look at the stability turning points B and C and deduce the stability behavior for any DM particle mass $m$. At the stability turning point B, as can be recognized in \cref{fig:mkms:rhop-10}, the core and the halo masses follow a simple power law, i.e. ${\SYMmassDM}^\text{\tiny (B)} \sim \brackets{{M_K}^\text{\tiny (B)}}^\alpha$, while point C is at a nearly constant core mass ${M_K}^\text{\tiny (C)}={M_K}^\text{\tiny (max)}\simeq M_{\rm OV}$ independent of $\SYMmassDM$. When ${\SYMmassDM}^\text{\tiny (B)}$ reaches the maximal DM halo mass, i.e. ${\SYMmassDM}^\text{\tiny (B)} = {\SYMmassDM}^\text{\tiny (max)}$ (the top tip of the green triangle) then the minimum and maximum bounds on the core mass (points B and C) merge, i.e., ${M_K}^\text{\tiny (B)}={M_K}^\text{\tiny (max)}$. Therefore, we may construct the normalized relation
\begin{equation}
    \label{eqn:relation-mkms}
    \frac{{\SYMmassDM}^\text{\tiny (B)}}{{\SYMmassDM}^\text{\tiny (max)}} = \brackets{\frac{{M_K}^\text{\tiny (B)}}{{M_K}^\text{\tiny (max)}}}^{\alpha}.
\end{equation} We can fit this equation numerically to estimate the maximal DM halo mass ${\SYMmassDM}^\text{\tiny (max)}$ and the exponent $\alpha$ for a given particle mass $m$. The values are listed in \cref{tbl:coefficients}. Further, we find ${M_K}^\text{\tiny (max)} \approx 0.82 M\approx M_{\rm OV}$ and very roughly a power law between the mass scale $M$ and the maximal halo mass. Using \cref{eqn:scale:mass} to replace $M$ we obtain estimations for the maximal core and halo masses depending on the particle mass: \begin{align}
    \label{eqn:approx:max-core-mass}
    {M_K}^\text{\tiny (max)} &\approx \SI{6.3E7}{\Msun} \brackets{\frac{m c^2}{\SI{100}{\kilo\eV}}}^{-2},\\
    \label{eqn:approx:max-halo-mass}
    {\SYMmassDM}^\text{\tiny (max)} &\approx \SI{2.8E14}{\Msun} \brackets{\frac{m c^2}{\SI{100}{\kilo\eV}}}^{-4.6}.
\end{align} The value of the maximal halo mass given by \cref{eqn:approx:max-halo-mass} is overestimated for particle masses close to the maximal particle mass of $\SI{387}{\kilo\eV}$, where the value is about $\SI{35}{\percent}$ larger, though still giving a proper estimation of the order of magnitude.

Interestingly, from \cref{eqn:approx:max-core-mass,eqn:approx:max-halo-mass} we obtain a relation between the maximal core and maximal DM halo mass independent of the particle mass, \begin{equation}
    \label{eqn:core-halo-mass-relation}
    {\SYMmassDM}^\text{\tiny (max)} \approx \SI{2.1E10}{\Msun} \brackets{\frac{{M_K}^\text{\tiny (max)}}{\SI{E6}{\Msun}}}^{2.3}.
\end{equation}

\newcommand{\heading}[1]{\multicolumn{1}{c}{#1}}
\begin{table}
    \caption{Coefficients of \cref{eqn:relation-mkms} and estimation of halo mass for different particle masses. The values have been numerically extracted from the stability analysis of solutions constrained by $M_K = \SI{4.2E6}{\Msun}$ and $\rho_p = \SI{E-2}{\Msun\per\parsec\cubed}$. All values are rounded.}
    \centering
    \begin{tblr}{
        width = \hsize,
        colspec = {
            Q[1, si={table-format=3}]
            Q[1, si={table-format=1.2}]
            Q[1.5, si={table-format=1.1e1}]
            Q[1.5, si={table-format=1.1e1}]
            Q[1.6, si={table-format=1.1e1}]
        }
    }
        \toprule
        \text{$mc^2$} & $\alpha$ & $M$ & ${M_K}^\text{\tiny (max)}$ & ${\SYMmassDM}^\text{\tiny (max)}$\\
        $\textcolor{gray}{\si{\kilo\eV}}$ & & $\textcolor{gray}{\si{\Msun}}$ & $\textcolor{gray}{\si{\Msun}}$ & $\textcolor{gray}{\si{\Msun}}$ \\
        \midrule
        56  & 3.17 & 2.5E8 & 2.0E8 & 4.0E15 \\
        200 & 3.25 & 1.9E7 & 1.6E7 & 1.2E13 \\
        250 & 3.31 & 1.2E7 & 1.0E7 & 4.2E12 \\
        387 & 3.32 & 5.1E6 & 4.2E6 & 3.0E11 \\
        \bottomrule
    \end{tblr}
    \label{tbl:coefficients}
\end{table}

\begin{figure}%
	\centering%
	\includegraphics[width=\hsize]{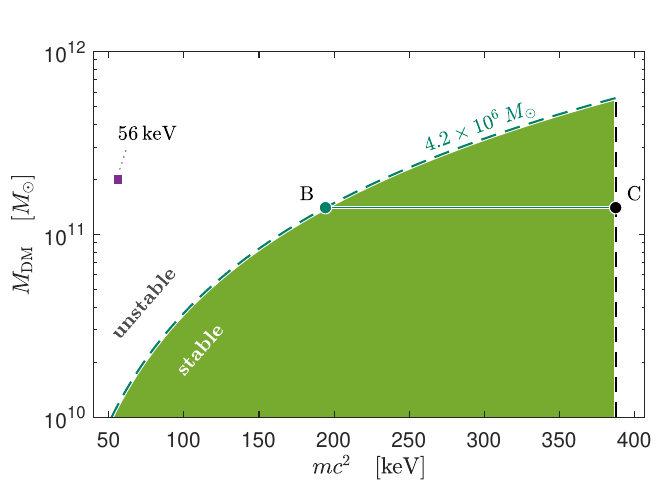}
	\caption{A simplified stability diagram as inferred numerically from \cref{fig:mkms:rhop-10}. It shows the relation between the DM particle mass $m$ and the DM halo mass $\SYMmassDM$ for stable \textit{core}--\textit{halo} solutions having a quantum core mass ${M_K}^\text{\tiny (MW)} = \SI{4.2E6}{\Msun}$ as shown by the (green-dashed) curve. The colored area represents possible options for stable configurations fulfilling the condition ${M_K}^\text{\tiny (MW)} \in [{M_K}^\text{\tiny (min)}, {M_K}^\text{\tiny (max)}]$. The numerical bounds for $M_K$ are given in \cref{eqn:approx:min-core-mass,eqn:approx:max-core-mass}.  Importantly, only DM halo masses $\SYMmassDM$ above $\SI{1.4E11}{\Msun}$ (i.e. limited from below by the line $B-C$) and up to the tip of the green-colored area at $\SI{5.5E11}{\Msun}$ are allowed by recent GAIA DR3 rotation curve data \citep{2023A&A...678A.208J}.}%
	\label{fig:particle-mass-bounds}%
\end{figure}

With the upper bounds given by \cref{eqn:approx:max-core-mass,eqn:approx:max-halo-mass} in \cref{eqn:relation-mkms} and for some average $\alpha = 3.3$ as a representative we obtain a relation for the minimal stable core mass ${M_K}^\text{\tiny (min)} \equiv {M_K}^\text{\tiny (B)}$ depending on the particle mass $m$ and the DM halo mass $\SYMmassDM \equiv {\SYMmassDM}^\text{\tiny (B)} $, \begin{equation}
    \label{eqn:approx:min-core-mass}
    {M_K}^\text{\tiny (min)} \approx \SI{5.7E6}{\Msun} \brackets{\frac{m c^2}{\SI{100}{\kilo\eV}}}^{-0.6} \brackets{\frac{\SYMmassDM}{\SI{E11}{\Msun}}}^{0.3}.
\end{equation}

With those inferred upper and lower bounds for the core and halo masses we may construct a simplified stability diagram in a $(mc^2, \SYMmassDM)$-plane, which allows us to estimate quickly the lower and upper bounds of the DM particle mass for any allowed average-size galaxy harboring a DM nucleus with the condition ${M_K}^\text{\tiny (MW)} \in [{M_K}^\text{\tiny (min)}, {M_K}^\text{\tiny (max)}]$.

In \cref{fig:particle-mass-bounds} we show such a stability diagram for ${M_K}^\text{\tiny (MW)} = \SI{4.2E6}{\Msun}$. The stable configurations are represented by a green area. That area is bounded from top-left by the minimal necessary core mass ${M_K}^\text{\tiny (min)} = {M_K}^\text{\tiny (MW)}$ (dark green dashed line) and from right by the maximal stable core mass ${M_K}^\text{\tiny (max)} = {M_K}^\text{\tiny (MW)}$ (dark dash line). These bounds correspond to the stability turning points B and C, respectively.

For a DM halo mass of ${\SYMmassDM}^\text{\tiny (MW)} = \SI{1.4E11}{\Msun}$ as considered in \cref{sec:results:comparison} we may bound the DM particle mass to the range $\SIrange{194}{387}{\kilo\eV}$. See horizontal dark green line in \cref{fig:particle-mass-bounds}.

While the upper bound of the DM particle mass is determined only by the core mass, see \cref{eqn:approx:max-core-mass}, the lower bound additionally depends on the DM halo mass, see \cref{eqn:approx:min-core-mass}. In general, the range of possible candidates for the particle mass increases for low DM halo masses and decreases for high DM halo masses. Moreover, notice that \cref{fig:particle-mass-bounds} exposes clearly a maximal DM halo mass which is given by \cref{eqn:core-halo-mass-relation}, i.e. ${\SYMmassDM}^\text{\tiny (MW)} \lesssim \SI{5.5E11}{\Msun}$. Interestingly, this theoretical upper limit coincides approximately with an independent upper limit of the total MW mass (baryons included) as observationally inferred in \cite{2023A&A...678A.208J}.

In previous works --- before this stability analysis --- a DM particle mass at the former lower bound of $\SI{56}{\kilo\eV}$ has been usually considered (e.g. \cite{2020A&A...641A..34B}). As illustrated in \cref{fig:particle-mass-bounds}, even if equilibrium fermionic profiles with ${\SYMmassDM}\approx \SI{E11}{\Msun}$ exist for such a light particle mass reproducing galactic observations, they turn out to be unstable.

Notice also that the particle mass range $\SIrange{194}{387}{\kilo\eV}$ that we have found from our stability analysis implies that the radius of the fermion ball is always much smaller than $\SI{6E-4}{\parsec}$, the S2 star pericenter, since $m c^2 > \SI{54.6}{\kilo\eV}$ (see \cref{sec:choice-param}). This is an interesting prediction of the statistical theory, as it does not rely on this observational constraint.

In \cref{sec_bounds} we give further explicit bounds on the mass of the fermionic DM particle. 

Nevertheless, we want to emphasize that the analysis has been done for a representative plateau density of $\rho_p = \SI{E-2}{\Msun\per\parsec\cubed}$, though we have verified that the stability analysis is not sensitive to slight variations of $\rho_p$. Hence, it is expected that the DM particle mass bounds, in particular the lower bound, will not change significantly.

\subsection{Implication on MW-like galaxies}
\label{sec:results:galaxy-implications}

\begin{figure}%
	\centering%
	\includegraphics[width=\hsize]{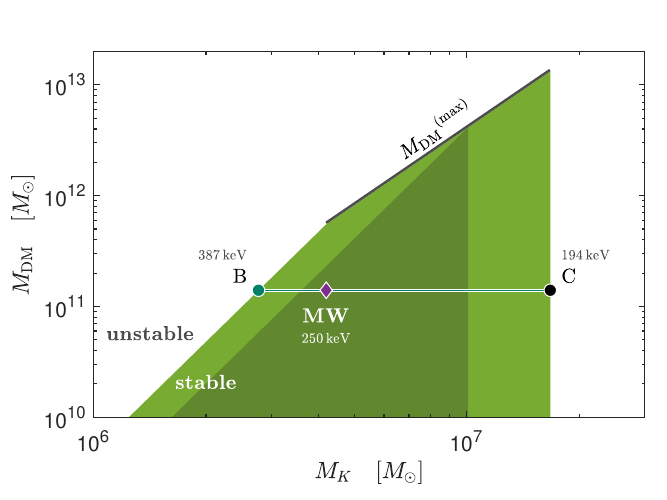}
	\caption{Stability diagram illustration for a particle mass within the range $mc^2 = \SIrange{194}{387}{\kilo\eV}$ as obtained in \cref{sec:results:mass-implications}. For a given particle mass  the set of stable solutions in the ($M_K$,$\SYMmassDM$) plane form a triangle and is illustrated for $mc^2 = \SI{250}{\kilo\eV}$ by a dark area. The whole set of stable solutions when limited to the mass bounds is represented by the green area. The minimal core mass (left bound) is given by \cref{eqn:approx:min-core-mass} for $mc^2 = \SI{387}{\kilo\eV}$ while the maximal core mass (right bound) is given by \cref{eqn:approx:max-core-mass} for $mc^2 = \SI{194}{\kilo\eV}$. Stable cores embedded in a DM halo of $\SYMmassDM$ = \SI{1.4E11}{\Msun} are indicated by the segment between the stability turning points B and C. Along the segment the MW from \cref{sec:results:comparison} is shown as a purple diamond. The DM halo mass $\SYMmassDM$ is bounded from top (solid line) following \cref{eqn:core-halo-mass-relation}, i.e. the whole set of all triangle tips each corresponding to a particle mass within the bounds. See main text for numerical values.}%
	\label{fig:stability-diagram:250}%
\end{figure}

We can use the results from \cref{sec:results:mass-implications} to make estimations for other (MW-like) galaxies with a DM halo mass of the order $\SI{E11}{\Msun}$ while the mass of the central compact object and the plateau density are unknown. We do not know the particle mass, either, but can at least use the above bounds from the MW analysis, i.e. $mc^2 = \SIrange{194}{387}{\kilo\eV}$. For every particle mass the set of stable solutions form a triangle in the ($M_K$, $\SYMmassDM$) plane as shown in \cref{fig:mkms:rhop-10}. The union of all stable sets corresponding to the obtained particle mass range is illustrated in \cref{fig:stability-diagram:250}.

We remind here that a DM plateau density of about $\SI{E-2}{\Msun\per\parsec\cubed}$ is only provided for a core mass $M_K = \SI{4.2E6}{\Msun}$. For the stable solutions along a caloric curve the plateau density may vary over some orders of magnitude. However, when we focus for a moment solely on the relation between the core mass $M_K$ and the DM halo mass $M_{\rm DM}$ we find interesting bounds for these galactic parameters.

The maximal core mass is given by \cref{eqn:approx:max-core-mass} for the minimal particle mass of $mc^2 = \SI{194}{\kilo\eV}$ and is independent of the DM halo mass. We obtain \begin{equation}
    \label{eqn:implications:min-core-mass}
    M_K \lesssim \SI{1.7E7}{\Msun}.
\end{equation}

The minimal core mass is given by \cref{eqn:approx:min-core-mass} for a particle mass of $mc^2 = \SI{387}{\kilo\eV}$ and depends additionally on the DM halo mass, \begin{equation}
    \label{eqn:implications:max-core-mass-a}
    M_K \gtrsim \SI{2.6E6}{\Msun} \brackets{\frac{\SYMmassDM}{\SI{E11}{\Msun}}}^{0.3}.
\end{equation} That lower bound holds only up to a maximal DM halo mass of about $\SI{5.5E11}{\Msun}$ as given by \cref{eqn:approx:max-halo-mass} for $\SI{387}{\kilo\eV}$. For more massive galaxies the particle mass must be reduced to fulfill the maximal DM halo mass relation. For that case the minimal core mass follows from \cref{eqn:core-halo-mass-relation} as \begin{equation}
    \label{eqn:implications:min-core-mass-b}
    M_K \gtrsim \SI{2.0E6}{\Msun} \brackets{\frac{\SYMmassDM}{\SI{E11}{\Msun}}}^{0.43}.
\end{equation} For a DM halo mass of $\SYMmassDM = \SI{1.4E11}{\Msun}$ as considered in \cref{sec:results:comparison} we find that embedded fermionic cores are stable when their mass is in the range $M_K \approx \SIrange{2.8E6}{1.7E7}{\Msun}$.

Finally, we note that from this analysis we obtain a maximal DM halo mass ${\SYMmassDM} \lesssim \SI{1.3E13}{\Msun}$ (as given by \cref{eqn:approx:max-halo-mass} for $\SI{194}{\kilo\eV}$) to be stable. Hence, a core mass in the proper range stabilizes the DM mass distribution. More massive galaxies with a quantum core-halo structure are necessary unstable regardless of the quantum core mass. In \cref{sec_bounds} we give further explicit bounds on the core and halo masses for a fiducial particle mass.

Although the bounds for core and halo mass look somewhat realistic --- i.e. with a particle mass of $mc^2 = \SI{250}{\kilo\eV}$ and for a galaxy with a DM halo mass of the order $\SI{E11}{\Msun}$ we obtain stable configurations only for core masses of the order $\SIrange{E6}{E7}{\Msun}$ --- these results must be taken with caution because the plateau density covers values over a wide range, about {\sisetup{print-zero-exponent = true} $\SIrange{E-6}{E0}{\Msun\per\parsec\cubed}$}. The lowest value is reached (simply speaking) for the minimal core mass while the highest value is reached for the maximal core mass. It is unclear if more massive galaxies, in particular in the range $\sim \SIrange{E11}{E13}{\Msun}$, with such extreme plateau densities exist. Probably even larger plateau densities may be reached when extrapolated to less massive galaxies ($M_{\rm DM} \lesssim \SI{E10}{\Msun}$).

\begin{figure}%
	\centering%
	\includegraphics[width=\hsize]{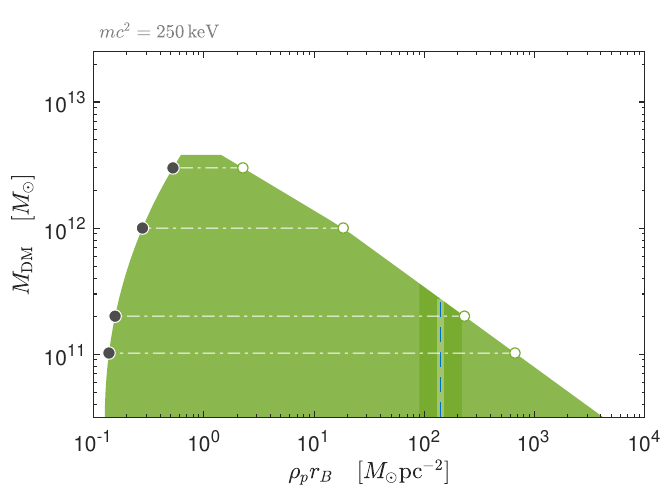}
	\caption{Stable configurations illustrated as a green area for the case of a particle mass of $\SI{250}{\kilo\eV}$ like in \cref{fig:mkms:rhop-10}. Shown are the surface density $\Sigma_{0} \approx \rho_p r_{\rm B}$ and the DM halo mass $M_{\rm DM}$. The blue vertical dashed line shows the observed mean surface density \cite{2009MNRAS.397.1169D}. The dark shaded region around the dashed line illustrates the $1\sigma$ uncertainty. The shown window of $\Sigma_0$ covers approximately the $3\sigma$ uncertainty.}%
	\label{fig:surface-density:250}%
\end{figure}

Despite the possibility of extreme $\rho_p$ values we find surprisingly a good match with surface density observations \cite{2009MNRAS.397.1169D} for a particle mass of $mc^2 = \SI{250}{\kilo\eV}$. The surface density is usually considered to be approximately constant for about 14 orders of magnitude in absolute magnitude ($M_B$), \begin{equation}
\label{eqn:Donato}
	\Sigma_{\rm 0D} = \rho_{\rm 0D} r_0 \approx 140_{-50}^{+80} \si{\Msun\per\parsec\squared},
\end{equation} with $\rho_{\rm 0D}$ being the \textit{central} DM halo density and $r_0$ the one-halo-scale-length --- both for the Burkert model. At $r_0$ the density falls to one-forth of the central density, i.e. $\rho(r_0) = \rho_{\rm 0D}/4$. Here, the \textit{center} in the Burkert model corresponds to the plateau in the fermionic DM model, i.e. $\rho_{\rm 0D} \approx \rho_{p}$. Following the definition of the Burkert radius $r_0$, we identify the one-halo-scale-length $r_B$ of the fermionic model such that $\rho(r_B) = \rho_p/4$. Hence, we estimate the surface density by the product $\Sigma_0 \approx \rho_p r_B$ for all solutions along a caloric curve. The results are illustrated in \cref{fig:surface-density:250}. All stable solutions (for $M_{\rm DM} \gtrsim \SI{5E10}{\Msun}$) cover very well the observed surface densities including the $3\sigma$ uncertainty, e.g. about $\SIrange{E-1}{E3}{\Msun\per\parsec\squared}$. See also \cite{2023ApJ...945....1K} for a similar application with similar results.

It is left for future research --- and out of scope of this paper --- how the stability behavior changes for order-of-magnitude variations on the $\rho_p$ value and whether any new bounds from other galaxies may be obtained. For instance, for large elliptical galaxies it is expected that the plateau density decreases to $\rho_p \sim \SI{E-3}{\Msun\per\parsec\cubed}$ while the core mass and DM halo mass increase in general \citep{2019PDU....24..278A}.


\section{Conclusions}
\label{sec:conclusion}

We have analyzed the stability of different, average-sized fermionic DM distributions with an embedded fermionic nucleus in its center, being a possible alternative to an SBMH, giving special attention to the Milky Way (in that case, Sgr A* would be a degenerate fermion ball rather than a BH). The results have important implications on the DM particle mass, on the possible relation between the DM quantum core and halo masses, on the maximal DM halo mass, and on the morphology of the DM halo including its outer halo tail.

In a former work, \cite{2018PDU....21...82A} found a family of fermionic solutions describing our Galaxy but without any stability analysis leaving the question open which solutions are stable. As shown in that work, the fermionic \textit{core}--\textit{halo} solutions depend on the particle mass which has been constrained by observables, in particular by the S-stars in the Galactic center. It was further demonstrated in \cite{2020A&A...641A..34B,2021MNRAS.505L..64B,2022MNRAS.511L..35A} that the astrometric data of the 17 best resolved S-cluster stars (including S-2 and the G-2 object) can be explained with a compact enough fermion-core centered in SgrA*, while further accounting for the relativistic redshift and relativistic periapsis precession effects measured for the S-2 star. More recently, and for particle masses corresponding to the stable branch of solutions analyzed here, in \cite{2024MNRAS.534.1217P} a fully relativistic ray-tracing technique was applied to Milky Way-like galaxies harboring a dense fermion-core surrounded by an accretion disk. It was shown that central brightness depression (i.e. shadow-like features) with sizes as observed by the Event Horizon Telescope \cite{2022ApJ...930L..13E} are possible for compact enough fermion cores.

Further support to the fermion-ball alternative has been recently given for other galaxy types of average size, including the case of particle masses falling within the stability region. Indeed, for low luminosity active-like galaxies, in \cite{2024A&A...685A..24M} it was demonstrated that (i) there is always a given fermion-core compactness that produces a luminosity spectrum that is almost indistinguishable from that of a Schwarzschild BH of the same mass as the fermionic DM core; and (ii) the accretion disk can enter deep inside the non-rotating fermionic DM core, allowing accretion-powered efficiencies as high as $\SI{28}{\percent}$, which is comparable to that of a highly rotating Kerr BH. This shows that the fermionic model opens new avenues of research for two seemingly unrelated topics: low mass AGN and DM physics.

In the case of the Milky Way, and extending the former work \citep{2018PDU....21...82A} with a proper stability analysis we find new bounds for the particle mass of about $mc^2 \approx \SIrange{194}{387}{\kilo\eV}$ able to explain the DM distribution in the Galaxy.

The lower particle mass bound is a consequence of imposing thermodynamical stability on the \textit{core}--\textit{halo} solutions having a quantum core with a mass of $\SI{4.2E6}{\Msun}$ being an alternative to the black hole hypothesis at the Galaxy center. It is determined by the stability change from unstable to stable core-halo solutions (labeled as point B in \cref{fig:caloric-curves:250}). Its precise value depends on the DM halo mass and increases for more massive halos (see \cref{sec:results:mass-implications} and \cref{sec_bounds}).

The upper bound is constrained by the core mass identified with $M_K$ whose maximum (labeled as point C in \cref{fig:caloric-curves:250}) is given approximately by the OV limit of gravitational collapse into a black hole. That is, the larger the particle mass $m$ the lower the maximum of $M_K$ (see for instance the bottom panel of \cref{fig:mkms:rhop-10}). Therefore, for a particle mass $mc^2 \gtrsim \SI{387}{\kilo\eV}$ the maximum of $M_K$ falls below the observationally constrained core mass of $\SI{4.2E6}{\Msun}$ such that there are no solutions fulfilling the $M_K$ constraint. In that case, the quantum core becomes unstable and is expected to collapse towards a supermassive black hole.

For a better understanding we encoded the stability behavior in a core-halo mass diagram, i.e. $M_K$ vs. $\SYMmassDM$ as in \cref{fig:mkms:rhop-10}, where the stable solutions are represented by a bounded triangle-shape area. Depending on the particle mass and other galaxy constraints, such as plateau density and core mass, the resulting stable area covers different ranges in $M_K$ and $\SYMmassDM$. The properties of these stable areas and especially their relation to the upper and lower bounds of the particle mass have been extracted numerically to construct a more concise stability diagram as shown in \cref{fig:particle-mass-bounds} for the MW analysis.

When stable fermionic solutions --- for a fiducial particle mass $mc^2=\SI{250}{\kilo\eV}$ --- together with standard baryonic components are asked to fulfill the rotation curve of the Galaxy, we find an excellent agreement with the most recent GAIA DR3 rotation curve data \citep{2023A&A...678A.208J}. We considered a DM halo mass of $\SYMmassDM=\SI{1.4E11}{\Msun}$ being the dominant mass component of our Galaxy and find a total mass of about \SI{2.0E11}{\Msun} including baryons. For the same stable solution we obtain a local DM halo density of about $\SI{0.4}{\giga\eV \speedoflight\squared \per \centi\metre}$ as independently inferred in \cite{2021A&A...653A..86W}.

A novel result of our study is the existence of a maximum DM halo mass ${\SYMmassDM}^\text{\tiny (max)}$ such that above that maximum there are no stable core-halo solutions. This maximum mass is reached when the solutions B and C of the caloric curve merge so that the stable core-halo branch disappears above such a limiting mass. For our Galaxy with a known value for the mass of the central dark compact object ${M_K}^\text{\tiny (MW)}=\SI{4.2E6}{\Msun}$, associated with SgrA*, we obtain from \cref{eqn:core-halo-mass-relation} an uppermost bound for the DM halo mass of about $\SI{5.5E11}{\Msun}$ interestingly in line with the independent upper limit inferred from observations \citep{2023A&A...678A.208J}.

More generally, for other (MW-like) galaxies, using \cref{eqn:approx:max-halo-mass} with the allowed range of the particle mass $mc^2 \approx \SIrange{194}{387}{\kilo\eV}$, we find that their DM halo mass must be below an upper bound of about $\SI{1.3E13}{\Msun}$ to be stable. Below that maximum mass galaxies similar to the MW should harbor a fermion ball with a core mass between $\SIrange{E6}{E7}{\Msun}$ (see \cref{sec:results:galaxy-implications}). By contrast, when extrapolated to other types of galaxies, these results suggest that sufficiently large DM halos should contain a supermassive black hole rather than a fermion ball (see \cref{sec:results:galaxy-implications} and \cref{sec_bounds}).
Following our results the transition should take place at a halo mass of the order $\SI{E12}{\Msun}$ depending on the DM particle mass.

This argument remains, however, qualitative because in this work the caloric curves have been constructed for MW-like galaxies, having for example a DM plateau density $\rho_p = \SI{1.0E-2}{\Msun\per\parsec\cubed}$ as obtained from the work of \cite{2018PDU....21...82A}. These values change for other galaxies, especially with larger and more massive DM halos \citep{2019PDU....24..278A}. Whether those other galaxies embedding a fermionic nucleus are stable remains an open question for further works.

We compared our results against the observed surface density relation \cite{2009MNRAS.397.1169D}. For a fiducial particle mass of $\SI{250}{\kilo\eV}$ and for galaxies with $M_{\rm DM} \gtrsim \SI{5E10}{\Msun}$ we find that all stable solutions are well covered by the $3\sigma$ uncertainty of $\Sigma_{0D}$. A better constrained value of the surface density would allow us to put stronger constraints on the maximal DM halo mass, see \cref{fig:surface-density:250}. For other particle masses within the allowed range, especially for the lower ($\SI{194}{\kilo\eV}$) and upper bounds ($\SI{387}{\kilo\eV}$), stable solutions are less well covered by the $3\sigma$ uncertainty of $\Sigma_{0D}$. Together with the application to our Galaxy in \cref{sec:results:comparison} these findings favor a particle mass of about $\SI{250}{\kilo\eV}$.

We also find that MW-like fermionic halos become stable only after a significant amount of particle evaporation from the outer halo. Such stable halos are polytropic with index $\nicefrac{5}{2}$ characterized by $W_p \ll 1$. A polytropic envelope can account for the latest GAIA DR3 results which show that beyond $\SI{25}{\kilo\parsec}$ the circular velocity decreases more rapidly than previously thought, hence that the DM halo of our Galaxy has a relatively strong mass concentration. In particular, these observational results strongly disfavor the traditional $\rho \sim r^{-3}$ Navarro-Frenk-White (NFW) profile as already shown in \cite{2024MNRAS.528..693O}, thus now favoring (stable) fermionic polytropic-like profiles in addition to typically assumed Einasto profiles \citep{2024MNRAS.528..693O}.

With a speculative extrapolation to much lower DM halo masses (e.g. dwarf galaxies) we expect an increase of the upper bound of the cutoff parameter $W_p$ such that a milder evaporation might be sufficient to stabilize a small galaxy, eventually with a power law-like halo tail in combination with core masses on the scale of intermediate black holes. Indeed in the case of small dwarf galaxies with $M_{\rm DM} \sim \SI{E8}{\Msun}$, we obtain from an equation like \cref{eqn:approx:min-core-mass} --- relating the minimal stable core mass with the DM halo mass --- a minimal core mass of about $\SI{E4}{\Msun}$ (as e.g. for a typical fermions mass of $mc^2 = \SI{250}{\kilo\eV}$).

\begin{figure}%
	\centering%
	\includegraphics[width=\hsize]{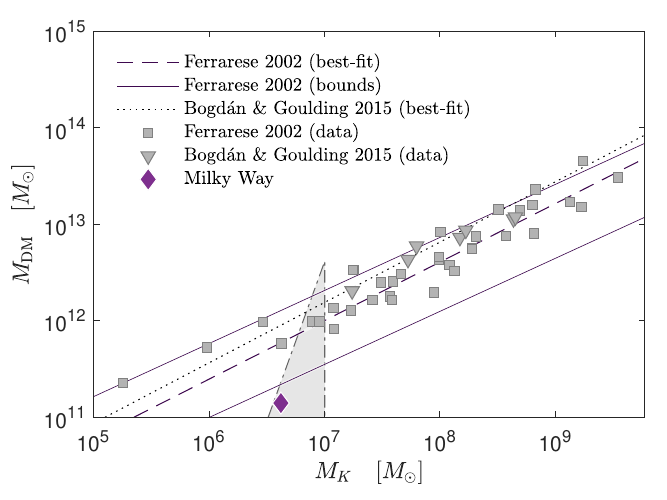}
	\caption{Comparison of the $M_{\rm BH}$-$M_{\rm DM}$ relation with the stable region of a fermionic DM distribution as determined for MW-like galaxies in \cref{sec:results}, with an exemplary DM particle mass of $mc^2 = \SI{250}{\kilo\eV}$. The MW from section \ref{sec:results:comparison} with $M_{\rm DM} = \SI{1.4E11}{\Msun}$ falls marginally below the Ferrarese strip since the DM halo mass has been updated in order to agree with recent GAIA DR3 analysis, which is considerably smaller than previous inferences. Central compact objects within the grey region may be fermion balls whereas central compact objects with mass $M_K > \SI{1E7}{\Msun}$ are necessarily black holes. Above that limit fermion balls with $mc^2 = \SI{250}{\kilo\eV}$ are unstable with respect to general relativity.}%
	\label{fig:ferrarese-comparison-250}%
\end{figure}

Remarkably for fermionic profiles aimed to model MW-like halos, the stable areas displayed in \cref{fig:mkms:rhop-10} (see also \cref{fig:ferrarese-comparison-250}) are always in agreement with the total mass vs central object mass relation of Ferrarese \cite{2002ApJ...578...90F} where the central supermassive object is now identified as the fermionic quantum core of mass $M_K$ (i.e. assumed to be a fermion ball). This suggests that the galaxies falling on the stable region (including our own galaxy indicated by a diamond in \cref{fig:ferrarese-comparison-250}) may harbor a fermion ball instead of a supermassive black hole as advocated in the present paper. However, we note that for larger galaxies of mass $\gtrsim \SI{E12}{\Msun}$, which constitutes most part of the Ferrarese relation, the mass of the central compact object is larger than the maximum mass $M_K \approx \SI{1.0E7}{\Msun}$ of a fermion ball made of particles with fiducial mass $mc^2 \approx \SI{250}{\kilo\eV}$. Thus, if DM is made of a single fermion species, the compact object harbored by these large galaxies cannot be a fermion ball (it would be unstable) but rather a supermassive black hole which can further grow to become large enough in sub-Giga year time-scales \citep{2023MNRAS.523.2209A,2024ApJ...961L..10A}. This transition at 
$\gtrsim \SI{E12}{\Msun}$ is nicely consistent with the maximum stable mass estimated from our theory. Thus, this fermionic theory for DM halos is still consistent with the presence of supermassive black holes (quasars) in active galactic nuclei (AGNs) like the one recently photographed in M87 ($M_h \sim \SI{E13}{\Msun}$ and  $M_{\rm BH} \sim \SI{E9}{\Msun}$). These results give a strong theoretical support to the idea that the fermion ball should become a black hole for large enough halo masses as conjectured in our previous papers \citep{2020EPJB...93..208A,2021MNRAS.502.4227A,2023MNRAS.523.2209A,2024ApJ...961L..10A}.

In future works, following the methodology described in this paper, we will consider other types of galaxies, ranging from small dwarf spheroidals to large ellipticals, and study their thermodynamical stability by varying the halo mass, plateau density, and DM particle mass. Of great interest are also galaxies with a DM halo mass close to the maximal DM halo because those galaxies would provide more stringent bounds on the DM particle mass. This ambitious research plan aims to seek up to which extent this fermionic theory for stable DM halos can predict the smallest up to the largest halo masses as observed in the galactic zoo, with their corresponding dark central objects.


\begin{acknowledgments}
    This work was founded by the Consejo Nacional de Investigaciones Científicas y Técnicas (CONICET), grant number 11220200102876CO.
\end{acknowledgments}


\appendix
\section{Maximum entropy principle}
\label{sec_grf}

We can obtain the equations of the fermionic mass distribution (see \cref{sec:theory}) all at once from the MEP. In the following we briefly review the formalism of this thermodynamic approach. We refer to \cite{2020EPJP..135..290C} for the details of the derivation --- albeit with a different set of variables which can be recovered by simple substitutions. Then, we discuss the physical justification of this maximization problem.



\subsection{Thermodynamical formalism}
\label{sec_os_formal}

We consider a general entropy functional of the form \begin{equation}
    \label{n76}
    \frac{S_{\rm tot}}{N k_{\rm B}} = \int_0^{r} \frac{s(r)}{\sigma} \brackets{1 - \frac{R}{r}\frac{M(r)}{M}}^{-1/2} \frac{r^2}{R^2} \frac{\d{r}}{R}
\end{equation} with the entropy density \begin{equation}
    \label{qgd9}
    s(r) = - k_{\rm B} \int C(f) \d{\vec p},
\end{equation} where $C(f)$ is a convex function (i.e. satisfying $C''(f) > 0$) and $\sigma = N k_{\rm B}/(4\pi R^3)$ is a scale factor. For systems with long-range interactions, such as self-gravitating systems, the mean field approximation is exact in a proper thermodynamic limit \cite{2013A&A...556A..93C}. In the microcanonical ensemble (for an isolated system) the statistical equilibrium state is obtained by maximizing the entropy $S_{\rm tot}$ at fixed total particle number $N_{\rm tot}$ and total energy $E_{\rm tot}$: \begin{equation}
    \label{maxent}
    \max \braces{S_{\rm tot} \; | \; N_{\rm tot}, E_{\rm tot} \; {\rm fixed}}.
\end{equation} The total particle number $N_{\rm tot}$ and the total energy $E_{\rm tot}$ are conserved --- or approximately conserved --- by the dynamics. The total mass-energy is $E_{\rm tot} = M_{\rm tot} c^2$.

To solve  \cref{maxent} we can proceed in two steps.

For the first step, at every position $r$ we maximize the entropy density $s(r)$ at fixed energy density $\rho(r)c^2$ and particle number density $n(r)$ with respect to variations on $f$. We write the variational problem for the first variations (extremization) under the form \begin{equation}
    \label{n80}
    \frac{\delta s}{\sigma} - \frac{1}{\beta(r)}\frac{\delta\rho}{\rho} + \frac{\alpha(r)}{\beta(r)}\frac{\delta n}{n} = 0,
\end{equation} where $\beta(r) = k_B T(r) / (mc^2)$, representing the local temperature, and $\alpha(r) = 1 + \mu(r) / (mc^2)$, representing the local chemical potential with rest-mass included, are local Lagrange multipliers. Using $s=-k_B\int C(f)\, \d{\vec p}$ together with $n=\int f\, \d{\vec p}$ and $\rho=\int f\epsilon\, \d{\vec p}$, this variational principle determines a distribution function of the form \citep{2020EPJP..135..290C} \begin{equation}
    \label{qgd11}
    f(r,\epsilon) = F\brackets{\frac{\epsilon - \alpha(r)}{\beta(r)}},
\end{equation} where $F(x) = (C')^{ - 1}( - x)$ is a monotonically decreasing function. Since $\delta^2 s = - k_B\int C''(f)\frac{(\delta f)^2}{2}\, \d{\vec p} < 0$ this distribution function is the global maximum of the entropy density at fixed energy density and particle number density. This corresponds to the condition of local thermodynamical equilibrium.


Substituting \cref{qgd11} into \cref{qgd9} we obtain after some calculations the Gibbs-Duhem relation \citep{2020EPJP..135..290C} \begin{eqnarray}
    \label{gd}
    \frac{s(r)}{\sigma} = \frac{1}{\beta(r)}\brackets{\frac{\rho(r)}{\rho} + \frac{P(r)}{\rho c^2} - \alpha(r) \frac{n(r)}{n}}.
\end{eqnarray} With this relation we can express the entropy $S_{\rm tot}$ as a functional of the local hydrodynamic variables $\rho(r)$ and $n(r)$.

For the second step, we maximize the entropy $S_{\rm tot}$ at fixed energy $E_{\rm tot}$ and particle number $N_{\rm tot}$ with respect to variations on $\rho(r)$ and $n(r)$. We write the variational problem for the first variations (extremization) under the form \begin{equation}
    \label{ax9}
    \frac{\delta S_{\rm tot}}{k_B} - \frac{1}{\beta_{\infty}}\frac{\delta E_{\rm tot}}{Mc^2} + \frac{\alpha_\infty}{\beta_{\infty}}\frac{\delta N_{\rm tot}}{N} = 0,
\end{equation} where $\beta_{\infty} = k_B T_{\infty}/(mc^2)$, representing the global temperature, and $\alpha_\infty = 1 + \mu_{\infty}/(mc^2)$, representing the global chemical potential with rest-mass included, are global Lagrange multipliers associated with the conservation of $E_{\rm tot}$ and $N_{\rm tot}$.


As detailed in \cite{2020EPJP..135..290C} the variational principle \cref{ax9} leads to the TOV \cref{eqn:mass,eqn:metric-potential} expressing the condition of hydrostatic equilibrium in general relativity and to  the Tolman and Klein relations \citep{1930PhRv...36.1791T,1949RvMP...21..531K}\begin{equation}
  \label{b103}
  \beta(r) = \beta_{\infty} \e^{-\nu(r)},
\end{equation} \begin{equation}
  \label{b104}
  \alpha(r) = \alpha_{\infty} \e^{-\nu(r)},
\end{equation} determining the spatial evolution of the temperature and chemical potential via the metric potential $\nu(r)$ with $g_{00} = \e^{2 \nu(r)}$. Here, $\beta_{\infty}$ and $\alpha_{\infty}$ represent the temperature and the chemical potential measured by an observer at infinity. We note that the ratio  $\alpha(r)/\beta(r) = \alpha_{\infty} / \beta_{\infty}$ is constant (uniform) since the temperature and the chemical potential are red-shifted in the same manner. As a result, the distribution function at statistical equilibrium is given by \begin{equation}
  \label{qgd15}
  f(r,\epsilon) = F\brackets{\frac{\epsilon}{\beta(r)} - \frac{\alpha_\infty}{\beta_\infty}}.
\end{equation}
This DF is a function $f=f({\cal E})$ of the energy at infinity ${\cal E}\equiv \epsilon e^{\nu(r)}$. Consequently, according to the relativistic Jeans theorem, it is a particular steady state of the Vlasov-Einstein equations. Furthermore, it is a monotonically decreasing function of the energy at infinity, $f'({\cal E})<0$, assuming positive temperatures ($\beta_\infty > 0$), which is the usual case.

We note that an extremum of entropy at fixed energy and particle number is necessarily isotropic (i.e. it does not depend on the angular momentum at infinity). As a result, the gas corresponding to the distribution function (\ref{qgd15}) is described by a barotropic equation of state $P(r) = P[\rho(r),\alpha_\infty/\beta_{\infty}]$, where the function $P[\rho,\alpha_{\infty}/\beta_{\infty}]$ is determined by the function $C(f)$ characterizing the entropy.

\subsection{Physical interpretation}
\label{sec_os_interpretation}

In thermodynamics the entropy $S=k_B \ln W$ is a measure of the disorder. It counts the number of microstates $W$ (complexions) corresponding to a given macrostate. As a result, the MEP determines the most probable distribution of particles at statistical equilibrium, i.e., the macrostate that is the most represented at the microscopic level.

For an ordinary gas this statistical equilibrium state results from a collisional relaxation due to strong encounters in the sense of Maxwell and Boltzmann. For systems with long-range interactions (like plasmas and self-gravitating systems) it results from a collisional relaxation due to numerous weak encounters in the sense of Landau and Chandrasekhar (see, e.g., \cite{2013A&A...556A..93C} for a review on kinetic theory).

The DF $f({\bf r},{\bf p},t)$ is governed by a kinetic equation --- e.g. the Boltzmann equation for an ordinary gas and the Landau or the Lenard-Balescu equation for systems with long-range interactions --- which satisfies an $H$\nobreakdash-theorem for the Boltzmann entropy: $\dot S_{\rm tot}[f]\ge 0$ \citep{2013A&A...556A..93C}. As a result, the system is expected to relax towards a maximum of entropy $S_{\rm tot}$ at fixed particle number $N_{\rm tot}$ and energy $E_{\rm tot}$. The optimization problem (\ref{maxent}) is a necessary and sufficient condition of thermodynamical stability. There is a specific difficulty for self-gravitating systems in the sense that there is no statistical equilibrium state (no entropy maximum) in an unbounded domain because of evaporation (see below). On the other hand, the relaxation time due to gravitational encounters is extremely long (especially in the case of fermions) and exceeds the age of the universe by many orders of magnitude. Therefore, the evolution of self-gravitating systems is essentially collisionless. Though, we notice the system could have a  collisional evolution with a much shorter relaxation time if the particles have a self-interaction of nongravitational origin as discussed in \cite{2022PhRvD.106d3538C}.

Let us consider a collisionless self-gravitating system described by the Vlasov equation. In that case, there is no direct relation to thermodynamics since there is no $H$\nobreakdash-theorem (any ``entropy'' $S_{\rm tot}[f]$ is conserved by the Vlasov equation). A collisionless system with long-range interactions can nevertheless reach a form of statistical equilibrium state from a process of violent relaxation on the coarse-grained scale in the sense of Lynden-Bell \cite{1967MNRAS.136..101L}. The coarse-grained DF $\overline{f}({\bf r},{\bf p},t)$  is governed by a kinetic equation (a generalized Landau or Lenard-Balescu equation) which satisfies an $H$\nobreakdash-theorem for the Lynden-Bell entropy: $\dot S_{\rm tot}[\overline{f}]\ge 0$. At equilibrium the coarse-grained DF is determined by a MEP in the sense of Lynden-Bell \cite{1967MNRAS.136..101L}; see \cite{2022PhyA..60628089C} for a recent review on the theory of violent relaxation for systems with long-range interactions. However, for self-gravitating systems there is no Lynden-Bell maximum entropy state in an unbounded domain. Therefore, the quasistationary state reached by the system as a result of violent relaxation is a stable stationary state of the Vlasov equation resulting from an incomplete violent relaxation and/or being affected by particle evaporation effects \cite{1967MNRAS.136..101L,1998MNRAS.300..981C}.

So far, we have discussed the statistical equilibrium state and the thermodynamical stability of collisional and coarse-grained collisionless systems with long-range interactions. Let us now consider the dynamical stability of collisionless systems described by the Vlasov equation. 

The extremization of a functional $S_{\rm tot}$ of the form (\ref{n76}) at fixed energy and particle number determines a particular steady state of the Vlasov equation of the form $f=f({\cal E})$ with $f'({\cal E})<0$. We note that $S_{\rm tot}$ is not a thermodynamical entropy in that case since there is no $H$\nobreakdash-theorem. It is just a particular Casimir (the Vlasov equation conserves an infinity of Casimirs). 

One can show in full generality that the maximization problem (\ref{maxent}) provides a sufficient condition of formal non-linear dynamical stability. We thus conclude that thermodynamical stability -- when a statistical equilibrium state exists -- implies dynamical stability \citep{2010JSMTE..06..001C}.

Furthermore, there is a conjecture by Ipser \cite{1980ApJ...238.1101I} that, in general relativity, the maximization problem (\ref{maxent}) provides a necessary and sufficient condition of dynamical stability with respect to the Vlasov-Einstein equations. If the conjecture of Ipser \cite{1980ApJ...238.1101I} is correct, then thermodynamical stability -- in the restricted sense given below -- would be equivalent to dynamical stability in general relativity.

By contrast, in Newtonian gravity, the maximization problem (\ref{maxent}) just provides a sufficient condition of dynamical stability with respect to the Vlasov-Poisson equations. Actually, it can be shown that all the DFs of the form $f = f(\epsilon)$ with $f'(\epsilon) < 0$ are dynamically stable with respect to the Vlasov-Poisson equations even those that do not maximize the functional $S_{\rm tot}$ at fixed particle number and energy \citep{Binney2008}. This is a distinctive difference between Newtonian gravity and general relativity.

Let us now try to combine these notions of dynamics and thermodynamics with a physical intuition of the evolution of self-gravitating systems. 

For fermions the thermodynamical entropy is the Fermi-Dirac entropy \begin{equation}
    \frac{s(r)}{k_{\rm B}} = -\eta_0\int \brackets[\frac{\eta_0}{A}]{f \ln f + \parantheses{1 - f} \ln \parantheses{1 - f}}\, \d{\vec p},
\end{equation} where $\eta_0 = g / h^3 $ is the maximum value of the DF fixed by the Pauli exclusion principle. The equilibrium distribution is the Fermi-Dirac DF \begin{equation}
    f(r,\epsilon) = \frac{1}{1 + \e^{[\epsilon - \alpha(r)]/\beta(r)}}.
\end{equation} 
As explained above it represents either the true equilibrium state resulting from a collisional evolution or the Lynden-Bell equilibrium state on the coarse-grained scale resulting from a collisionless evolution (with the replacement $f\rightarrow \overline{f}$). However, when coupled to gravity the Fermi-Dirac DF leads to configurations with an infinite mass. Physically, there is no maximum entropy state because the system has the tendency to evaporate. 

Developing the kinetic theory of self-gravitating fermions by taking tidal effects into account,  it is possible to justify \citep{1998MNRAS.300..981C} the truncated Fermi-Dirac DF 
\begin{equation}
    \label{rfkingannexe}
    f(r,\epsilon) = \frac{1 - \e^{\brackets{\epsilon - \varepsilon(r)}/\beta(r)}}{1 + \e^{\brackets{\epsilon - \alpha(r)}/\beta(r)}},
\end{equation} 
where the escape energy is given by \begin{equation}
    \varepsilon(r) = \alpha(r) + \beta(r)\ln\parantheses{\frac{\eta_0}{A}}.
\end{equation} This  is a steady state of the Vlasov equation of the form of \cref{qgd15} called the fermionic King model \citep{2015PhRvD..92l3527C}. As a result, it extremizes a particular ``entropy'' of the form of \cref{n76} at fixed particle number and energy. This generalized entropy reads \citep{2015PhRvD..92l3527C} \begin{equation}\begin{split}
    \frac{s(r)}{k_{\rm B}} = & -A \int \brackets{\parantheses{1 + \frac{\eta_0}{A}f}\ln \parantheses{1 + \frac{\eta_0}{A}f} - \frac{\eta_0}{A}f} \,d{\vec p}\\
    & + \eta_0 \int \brackets[\frac{\eta_0}{A}]{\parantheses{1 - f}\ln \parantheses{1 - f} + f} \,\d{\vec p} \\
    & - \eta_0 \int \ln\parantheses{\frac{\eta_0}{A}} f \,\d{\vec p}.
    \label{rfkings}
\end{split}\end{equation}
Indeed, calculating $\delta s$ from Eq. (\ref{rfkings}) and substituting this expression into the variational principle from Eq. (\ref{n80}), or directly using Eq. (\ref{qgd11}), we obtain the truncated Fermi-Dirac DF from Eq. (\ref{rfkingannexe}).

For evaporating (tidally truncated) systems the situation is complicated because we have to consider an out\nobreakdash-of\nobreakdash-equilibrium problem. Therefore, it is not quite clear if thermodynamical arguments rigorously apply to this situation. However, since evaporation is a slow process, the fermionic King distribution (\ref{rfkingannexe}) must be both a stable steady state of the Vlasov equation and, in some sense, a maximum entropy state for the entropy (\ref{rfkings}). We may expect that an ``entropy'' maximum will be stable --- from a dynamical and a ``thermodynamical'' point of view --- while an ``entropy'' minimum (or saddle point) will be unstable at least from a ``thermodynamical'' point of view -- i.e. it will not be the most probable state.

It is therefore relevant (but not totally rigorous) to connect the maximization problem (\ref{maxent}) with the entropy (\ref{rfkings}) to the stability of the fermionic King distribution. In that case, we can use the machinery of thermodynamics. In particular, the stability of the system can be investigated by plotting the caloric curve $\beta_{\infty}(M_{\rm tot})$ (series of equilibria) and using the Poincaré-Katz turning point criterion (see \citep{2015PhRvD..92l3527C,2020EPJB...93..208A,2021MNRAS.502.4227A} and  section \ref{sec:method} for details).

In conclusion, our thermodynamical approach relies on a sort of adiabatic approximation. We are assuming that evaporation is a slow process so that the mass and the energy of the system are approximately conserved. More precisely, we assume that the system evolves through a succession of quasistationary states of the Vlasov equation whose mass and energy (and other internal parameters) slowly change with time because of evaporation. In the present context, the system follows a sequence of fermionic King distributions along the series of equilibria $\beta_{\infty}(M_{\rm tot})$ with decreasing mass-energy and increasing central density. This is what happens to globular clusters through collisional relaxation and probably also what happens to fermionic DM halos (at least in the framework of our model) whatever the source of evolution (collisionless relaxation or self-interaction) \citep{2015PhRvD..91f3531C,2015PhRvD..92l3527C}. This general picture should not be fundamentally altered by general relativity since relativistic effects are only relevant to the physics of the high density core and its eventual collapse, while evaporation mainly involves the (diluted) outer halo. Numerical estimates of the evaporation rate are in principle possible within the generalized Landau equation derived in \cite{1998MNRAS.300..981C}, which is the appropriate kinetic equation for our fermionic system. It would require calculating the diffusion current in terms of the fermionic parameters of our theory when applied to real galactic halos, though this is out of the scope of the present paper and left for a future work.

\subsection{Limitations of the statistical theory}
\label{sec_limitation}

The Vlasov equation, which governs the dynamical evolution of collisionless self-gravitating systems, admits an infinite number of stable stationary solutions. If we restrict ourselves to spherically symmetric systems, their DF  may generally depend on the energy and on the modulus of the angular momentum. Predicting the DF that describes DM halos is difficult. Statistical mechanics makes a clear prediction and has the interest to rely on a solid theoretical basis. However, it has its own limitations. In particular, it assumes ergodicity and complete relaxation, which may not always be fulfilled in practice.

In the present paper, we have assumed that statistical mechanics can be applied to DM halos. This is our working hypothesis. This is certainly an important hypothesis to test in detail before considering more complicated possibilities. Statistical mechanics leads to the Fermi-Dirac distribution (or more precisely to the fermionic King model) which depends only on the energy and thus implies that DM halos at statistical equilibrium are spherically symmetric and isotropic. Other models of DM halos such as \cite{2015ApJ...811....2H,2022ApJ...937...67W} also assume that the DF depends only on the energy.

It is possible that real DM halos have a radially biased velocity distribution in case of incomplete relaxation, such as the one observed in elliptical galaxies.  However, this leads to more complicated models, in which the DF depends on energy and angular momentum. Predicting the anisotropic DF is not easy, and analyzing the stability of anisotropic DFs is even harder. This is why we did not consider this possibility here. In addition, there are no quantum numerical simulations of fermionic DM halos that give information about their DF.

Our paper shows that a statistical mechanics model based on the fermionic King DF gives a remarkable agreement with the observations (see \cref{fig:rotation-curve:250,fig:rotation-curve-zoom:250}). Therefore, the fermionic King model justified by statistical mechanics seems to be a relevant model. This is a good first step before considering more complex models including anisotropy.

Our model assumes that collisionless relaxation produces a high density contrast between the core and the halo such that the core can mimic a SMBH. In practice, a purely collisionless evolution does not produce a DM halo with a high density contrast because the collisionless relaxation, which is driven by the fluctuations of the gravitational field, stops before statistical equilibrium is reached (incomplete relaxation). Gravitational encounters of fermions will not solve this problem because they are inefficient even in the central region of the system. This difficulty is discussed in \cite{2022PhRvD.106d3538C}. This difficulty could be circumvented if the fermions are self-interacting (see Sec. XI and footnote 57 of \cite{2022PhRvD.106d3538C}). A self-interaction could induce a dynamical evolution of the DM halo (like gravitational encounters in a globular cluster) and trigger a gravothermal catastrophe at the critical energy corresponding to the first turning point of the caloric curve (point A in \cref{fig:caloric-curves:250}) allowing the system to reach high densities. Interestingly, at this point, the King profile is similar to the Burkert profile (see Fig. 1 in \cite{2022PhRvD.106d3538C}) and it will remain so because the structure of the envelope is unaffected by core collapse. In the case of fermionic DM halos, core collapse may be arrested by quantum degeneracy (Pauli's exclusion principle) and lead to the formation of a stable quantum core (fermion ball). However, it was demonstrated in \cite{2020EPJB...93..208A}, and confirmed in \cite{2021MNRAS.502.4227A}, that the branch of condensed core-halo solutions presents itself a turning point of energy of general relativity origin (point C in \cref{fig:caloric-curves:250}) at which the core becomes unstable and collapses towards a SMBH. This type of scenario, where a slow (secular) gravothermal catastrophe is followed by a fast gravitational collapse towards an SMBH on a few dynamical timescales, is considered by Balberg {\it et al.} \cite{2002ApJ...568..475B} in the context of self-interacting dark matter (SIDM) and discussed further in \cite{2015PhRvD..91f3531C,2015PhRvD..92l3527C,2020EPJP..135..310C,2021MNRAS.502.4227A,2022PhRvD.106d3538C,2023MNRAS.523.2209A} in relation to fermionic DM halos. A self-interaction could also facilitate the relaxation of the system towards the statistical equilibrium state (Fermi-Dirac-like) and avoid (or limit) the problem of incomplete collisionless relaxation mentioned above.

Our results are only based on the assumption that DM halos are described by the fermionic King DF. The fermionic King DF is clearly a fundamental DF to study in detail before considering more complicated models. Of course, it would be nice to show from direct numerical simulations of quantum (fermionic) DM that this DF emerges naturally, but this appears to be very difficult. Therefore, the alternative strategy is to assume this DF (which is predicted by statistical mechanics to be the most probable equilibrium state) from the start and see if it can account for the observations. Our paper makes for the first time a direct connection between the caloric curves of the self-gravitating Fermi gas (an important theoretical topic in itself with a long history \cite{2023mgm..conf.2230C}) and astrophysical data for the MW. Remarkably, we can explain the observations of the MW with the fermionic King model, analyze its thermodynamical stability, and constrain the range of the DM particle mass. Our study shows the importance of statistical mechanics to DM halos (when properly formulated) and its astrophysical applications.

\section{Range of quantum core mass values}
\label{sec:core-mass-extrema}
Specific core-halo solutions along a caloric curve show only a limited set of mass distributions. In particular, we are interested in the range of possible core masses $M_K$ and in the existence of a stable subset, both shown in \cref{fig:mkms:rhop-10}.

\begin{figure}%
	\centering%
	\includegraphics[width=\hsize]{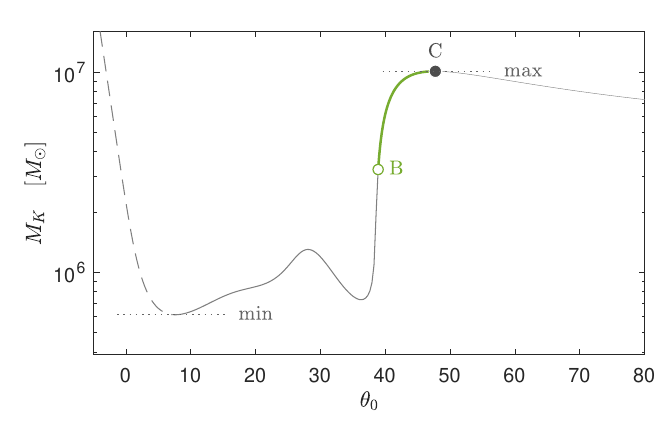}
	\caption{Illustration of a typical DM core mass profile as a function of the central fermion-degeneracy for solutions lying along the caloric curve of \cref{fig:caloric-curves:250} (upper panel). Continuous grey-thin lines represent unstable solutions. Green-thick line represents the stable solutions containing the  $M_K = \SI{4.2E6}{\Msun}$ quantum core mass as alternative to the BH scenario in SgrA*. Points B and C indicate a change of stability. The dashed curve represents the transition regime to the so called degenerate quantum cores with $\theta_0 \gtrsim 7$.}%
	\label{fig:core-mass-profile:250}%
\end{figure}

For a better understanding why the values for $M_K$ are bounded, in particular the lower bound ${M_K}^\text{\tiny (0)}$, we show in \cref{fig:core-mass-profile:250} a typical curve of the core mass $M_K$ vs. the central degeneracy parameter $\theta_0$. There we can see that a global minimum for $M_K$ exists at $\theta_0 \approx 7$ such that, as explained in \cref{sec:results}, for $\theta_0 \lesssim 7$ the mild degenerate cores do not fulfill the statistical quantum (fermionic) condition since the thermal de Broglie wavelength starts to fall below the inter-particle mean distance. In that case, there is no distinct nucleus.

Our main interest are core-halo solutions ($\theta \gtrsim 15$) with a clear nucleus, an extended plateau followed by a halo. In that core-halo regime the Keplerian mass suitably describes the dark compact object harbored in the galactic centers. Interestingly, in \cref{fig:core-mass-profile:250} we can identify a maximum of $M_K$ very close to the last stable point $C$. This maximum core mass at which the system becomes unstable practically coincides with the analytic expression of the OV limit shown in \cref{sec:choice-param}. As first demonstrated in \cite{2021MNRAS.502.4227A}, this maximum, when plotted in the central density ($\rho_0$) vs. total DM halo mass ($M_{\rm DM}$), leads to the old-known turning point instability criterion for gravitational collapse, which it is demonstrated to coincide with the last stable solution at point $C$ in the caloric curves. (See also Section 4 in \cite{2021MNRAS.502.4227A} for further details.)

It is worth mentioning that for larger DM halo masses $\SYMmassDM$, and at fixed $m$, the point $B$ moves to larger $\theta_0$ values while point $C$ remains nearly constant. Only when point $B$ approaches point $C$ then point $C$ moves to smaller $\theta_0$ values, though remains close to the maximum of \cref{fig:core-mass-profile:250}.

\section{Explicit bounds on the mass of the fermionic DM particle}
\label{sec_bounds}

From the numerical results used to make \cref{fig:mkms:rhop-10}, we find that the minimum stable core mass ${M_K}^\text{\tiny (B)}$ (corresponding to point B on the caloric curve of \cref{fig:caloric-curves:250}) depends on the DM halo mass $\SYMmassDM$ of the galaxy according to the scaling law
\begin{equation}
\label{mam}
\frac{\SYMmassDM}{M_{\odot}}=A(m)\left\lbrack\frac{{M_{K}}^\text{\tiny (B)}}{M_{\odot}}\right\rbrack^{\alpha(m)},
\end{equation}
where $\alpha(m)$ and $A(m)$ depend on the particle mass $m$. The exponent $\alpha\sim 3.3$ changes slowly with $m$ (see \cref{tbl:coefficients}) while the amplitude $A$ has a more significant dependence on $m$: $A(56)=1.8169\times 10^{-11}$, $A(200)=5.0935\times 10^{-11}$, $A(250)=2.7273\times 10^{-11}$ and $A(387)=3.2327\times 10^{-11}$.
On the other hand, the maximum stable core mass ${M_{K}}^\text{\tiny (max)}$, which coincides with the OV mass, is given by \cref{eqn:approx:max-core-mass}. The maximum halo mass ${\SYMmassDM}^\text{\tiny (max)}$ (top of the green triangle) above which there is no stable core mass is obtained by writing ${M_K}^\text{\tiny (B)}={M_{K}}^\text{\tiny (max)}$ in \cref{mam}, leading to the expression of ${\SYMmassDM}^\text{\tiny (max)}$ from \cref{eqn:approx:max-halo-mass}. We can then rewrite \cref{mam} under the form of \cref{eqn:relation-mkms}. The maximum mass ${\SYMmassDM}^\text{\tiny (max)}$ is reached when the points B and C in the caloric curve of \cref{fig:caloric-curves:250} merge so that the stable condensed branch disappears.

Using these results, we can obtain explicit bounds on the mass of the DM particle. Let us suppose that ${\SYMmassDM}^\text{\tiny (MW)}$ (mass of the MW) and ${M_{K}}^\text{\tiny (MW)}$ (quantum core mass of the MW identified with the dark compact object SgrA*) are known. We then have the following constraints:

(i) ${\SYMmassDM}^\text{\tiny (MW)} < {\SYMmassDM}^\text{\tiny (max)}$: this is a necessary condition to have a core-halo solution with a stable quantum core (note that the value of the quantum core mass is not prescribed in this criterion).

(ii) ${M_{K}}^\text{\tiny (MW)} < {M_K}^\text{\tiny (max)}$: the quantum core mass of the MW  must be smaller than the maximum mass ${M_K}^\text{\tiny (max)}$ (OV mass) in order to be stable in general relativity.

(iii) ${M_{K}}^\text{\tiny (MW)} > {M_K}^\text{\tiny (B)}$: the quantum core mass of the MW  must be larger than the minimum mass ${M_K}^\text{\tiny (B)}$ in order to be thermodynamically stable.

The goal now is to obtain explicit bounds on $m$ from the above constraints:

(i) Using \cref{eqn:approx:max-halo-mass} we obtain
\begin{equation}
    \frac{mc^2}{\SI{100}{\kilo\eV}} < 5.61 \brackets{\frac{{\SYMmassDM}^\text{\tiny (MW)}}{\SI{E11}{\Msun}}}^{-0.217}.  
\end{equation}
For the fiducial value ${\SYMmassDM}^\text{\tiny (MW)} = \SI{1.4E11}{\Msun}$ we get $mc^2 < \SI{521}{\kilo\eV}$. Above that value of the DM particle mass, the MW cannot support a stable quantum core of any mass. In that case, it should rather contain a SMBH.

(ii) A more stringent bound on $m$ is obtained from \cref{eqn:approx:max-core-mass}. We obtain
\begin{equation}
    \frac{mc^2}{\SI{100}{\kilo\eV}}< 7.94 \brackets{\frac{{M_{K}}^\text{\tiny (MW)}}{\SI{E6}{\Msun}}}^{-1/2}.  
\end{equation}
For the fiducial value ${M_{K}}^\text{\tiny (MW)} = \SI{4.2E6}{\Msun}$ we get $mc^2 < \SI{387}{\kilo\eV}$. This constraint is more restrictive (stronger) than the previous one but it relies on the knowledge of $M_{\rm K}$. Above that value of the DM particle mass, a quantum core of mass ${M_{K}}^\text{\tiny (MW)} = \SI{4.2E6}{\Msun}$ is unstable in general relativity and undergoes gravitational collapse. In that case, the dark compact object SgrA* should rather be identified with a SMBH.

(iii) Using \cref{eqn:approx:min-core-mass} we obtain
\begin{equation}
    \frac{mc^2}{\SI{100}{\kilo\eV}}> 17.5
    \brackets{\frac{{\SYMmassDM}^\text{\tiny (MW)}}{\SI{E11}{\Msun}}}^{1/2}
    \brackets{\frac{{M_{K}}^\text{\tiny (MW)}}{\SI{E6}{\Msun}}}^{-1.65}.  
\end{equation}
For the fiducial values ${\SYMmassDM}^\text{\tiny (MW)} = \SI{1.4E11}{\Msun}$ and ${M_{K}}^\text{\tiny (MW)} = \SI{4.2E6}{\Msun}$, we get $mc^2 > \SI{194}{\kilo\eV}$. Below that value of the DM particle mass, a quantum core of mass ${M_{K}}^\text{\tiny (MW)} = \SI{4.2E6}{\Msun}$ is thermodynamically unstable. 

The foregoing results give us explicit formulae determining the minimum and maximum bounds on the fermion mass $m$ as a function of the mass of the MW and the value of the dark compact object that it contains. This allows us to see how these bounds change by varying ${\SYMmassDM}^\text{\tiny (MW)}$ and ${M_{K}}^\text{\tiny (MW)}$.

We can also turn the arguments the other way round, i.e., we can fix the mass $m$ of the DM particle and obtain constraints on $M_K$ and $\SYMmassDM$ possibly valid for galaxies different from the MW (under the assumption that our adopted value of the plateau density remains approximately correct for such galaxies). In that case we find:  

(i)
\begin{equation}
\frac{\SYMmassDM}{\SI{E11}{\Msun}} < \num{2.8E3} \brackets{\frac{mc^2}{\SI{100}{\kilo\eV}}}^{-4.6}.
\end{equation}
For the fiducial value $mc^2=\SI{250}{\kilo\eV}$ we get $\SYMmassDM<\SI{4.14E12}{\Msun}$. Above that mass, there is no core-halo solution with a stable quantum core. Therefore, sufficiently large galaxies should contain a SMBH rather than a fermion ball.

(ii)
\begin{equation}
\frac{M_K}{\SI{E6}{\Msun}}<63\, \brackets{\frac{mc^2}{\SI{100}{\kilo\eV}}}^{-2}.
\end{equation}
For the fiducial value $mc^2=\SI{250}{\kilo\eV}$ we get $M_K<\SI{1.10E7}{\Msun}$. Above that maximum mass (OV limit), the quantum core becomes unstable in general relativity and undergoes gravitational collapse.

(iii)
\begin{equation}
\frac{M_K}{10^6\, M_{\odot}}>5.69\, \brackets{\frac{\SYMmassDM}{\SI{E11}{\Msun}}}^{0.303} \brackets{\frac{mc^2}{\SI{100}{\kilo\eV}}}^{-0.606}.
\end{equation}
This equation corresponds to the scaling relation from \cref{mam}. Under this form it is similar to a core mass -- halo mass relation. For the fiducial value $mc^2=\SI{250}{\kilo\eV}$ we get $M_K > \SI{3.27E6}{\Msun}\, ({\SYMmassDM}/\SI{E11}{\Msun})^{0.303}$. Below that minimum mass the quantum core becomes thermodynamically unstable.


\bibliography{references}
\end{document}